\documentclass[fleqn,10pt]{wlscirep}

\usepackage[utf8]{inputenc}
\usepackage[T1]{fontenc}

\usepackage{helvet}

\usepackage{sidecap}
\usepackage{float}
\usepackage{nccmath}
\usepackage{amssymb,amsmath}

\usepackage{caption}
\DeclareCaptionLabelFormat{myformat}{\textbf{Fig}~#2}
\title{
The role of excitation vector fields and all-polarisation state control of cavity magnonics}
\author[1]{Alban Joseph}
\author[2]{Jayakrishnan M. P. Nair}
\author[1]{Mawgan A. Smith}
\author[1]{Rory Holland}
\author[1]{Luke J. McLellan}
\author[3]{Isabella Boventer}
\author[4]{Tim Wolz}
\author[5]{Dmytro A. Bozhko}
\author[2]{Benedetta Flebus}
\author[1]{Martin P. Weides}
\author[1,*]{Rair Mac\^edo}
\affil[1]{James Watt School of Engineering, Electronics \& Nanoscale Engineering Division, University of Glasgow, Glasgow G12 8QQ, United Kingdom}
\affil[2]{Department of Physics,Boston College,140 Commonwealth Avenue,Chestnut Hill, MA 02467}
\affil[3]{Laboratoire Albert Fert, 1 Avenue Augustin Fresnel, 91767 Palaiseau, France}
\affil[4]{Institute of Physics, Karlsruhe Institute of Technology, 76131 Karlsruhe, Germany}
\affil[5]{Center for Magnetism and Magnetic Materials, Department of Physics and Energy Science, University of Colorado Colorado Springs, Colorado Springs, Colorado 80918, USA}
\affil[*]{Rair.Macedo@glasgow.ac.uk}

\begin{abstract}
Recently the field of cavity magnonics, a field focused on controlling the interaction between magnons and confined microwave photons within microwave resonators, has drawn significant attention as it offers a platform for enabling advancements in quantum- and spin-based technologies%~\cite{Li2020, lachance-quirionEntanglementbasedSingleshotDetection2020}
. 
Here, we introduce excitation vector fields, whose polarisation and profile can be easily tuned in a two-port cavity setup, thus acting as an effective experimental knob to explore the coupled dynamics of cavity magnon-polaritons. 
Moreover, we develop theoretical models that accurately predict and reproduce the experimental results for any polarisation state and field profile within the cavity resonator.
This versatile experimental platform offers a new avenue for controlling spin-photon interactions and as such also delivering a mechanism to readily control the exchange of information between hybrid systems.

\end{abstract}
\begin{document}

\flushbottom
\maketitle

\thispagestyle{empty}

\section*{Introduction}
$ $

Magnons, the quanta of spin waves in magnetic materials, have emerged as promising information carriers for developing novel classical and quantum technologies \cite{Kajiwara2010, Cornelissen2015, Jiang2023}. As a result, the field of cavity magnonics, which explores the strong coupling between magnons and photons confined within microwave cavities, has been the subject of intense research in recent years~\cite{ZareRameshti2022}. These cavity magnonic systems give rise to a quasiparticle known as the cavity-magnon polariton, and controlling the behaviour of this polariton in such systems, i.e. the hybridisation between cavity photons and magnons, offers a platform for manipulating and processing information carried by magnons. 
Notably, it has recently been demonstrated that magnons can couple to superconducting qubits through cavity photons and, as a consequence, cavity magnon-photon systems have garnered attention for their potential applications in quantum information processing \cite{tabuchiCoherentCouplingFerromagnetic2014, lachance-quirionEntanglementbasedSingleshotDetection2020, lachance_quirion19}.
Moreover, with the successful realisation and control of spin-based qubits using microwave photons, exploring and manipulating spin-photon interactions could potentially provide a powerful tool for accessing and controlling quantum states \cite{bhoiChapterTwoRoadmap2020}. Beyond quantum computing, microwave photons are commonly used to control magnonic technologies \cite{Chumak2015}, with applications ranging from spin-based signal processing \cite{Kostylev2005, Vogt2014, Khitun2012} to magnetic sensing \cite{Flower2019, flower19} and energy-efficient computing \cite{Chumak2014, Khitun2011}.
Therefore, further exploiting the interaction between magnons and photons offers a promising route to the controlling, manipulating, and discerning spin systems \cite{baityStrongMagnonPhoton2021}, contributing to advancements ranging from quantum technologies to magnon-based devices.

To be successful in these applications however, it is imperative to have a precise control over the coupling strength between the photons and magnons. The coupling strength dictates the rate of energy exchange, and hence controlling this rate would increase the flexibility of coherent information exchange. % \cite{wolz19}. 
This has resulted in extensive efforts dedicated to engineering and controlling the coupling through various %diverse methodologies
approaches\cite{zhang17, stenningMagneticControlMetamolecule2013}. Previous studies have explored  a range of methods, including the use of various types of resonators %strategies, involved a range of approaches
\cite{bhoiRobustMagnonphotonCoupling2017, bhoiStudyPhotonMagnon2014, raoLevelAttractionLevel2019, Kato2023}, modifications in materials \cite{Flower2019, boventerAntiferromagneticCavityMagnon2023, khivintsevNonlinearFerromagneticResonance2010a, Xu2024}, and adjustments in sample positioning within the cavity \cite{macedoElectromagneticApproachCavity2021, yangControlMagnonphotonLevel2019, raoLevelAttractionLevel2019, Wagle2024, Rao2023} to influence the coupling strength between magnons and photons. 
Additionally, researchers have developed tunable techniques that allow for transitioning between different coupling regimes without the need for physical modifications to the experimental setup \cite{boventer19b, zhang15b, yangControlMagnonphotonLevel2019, boventer19a}.  
Investigations into time-domain experiments, involving sequences of pulses to excite magnetic samples, have enabled dynamic control over coupling strength \cite{wolz19, Match2019, zhang15}, providing additional insights into the temporal aspects of magnon-photon interactions.
%engineer photons dissipation \cite{zhang17}. 

Up to now most works have neglected how the polarisation state of the excitation vector fields within the resonator can modify the coupling of the hybrid modes. Here, we demonstrate experimentally that by modifying the polarisation of cavity electromagnetic modes, one can efficiently switch from level-repulsion to coupling annihilation.
In doing so, we find that different polarisation states can be readily employed to tune the coupling strength.
%from strong coupling to complete decoupling. 
We show how a square cavity resonator -- excited by two ports -- can be used to generate multiple polarisation states, namely circularly, linearly and elliptically polarised modes, and these can dramatically modify the overall behaviour of the cavity-magnon hybridisation. For instance, we find that by using circularly polarised excitation fields, we can transition from complete decoupling to an enhancement in the coupling strength by a factor of $\sqrt{2}$ in comparison to excitation by a linearly polarised field. 
We provide two theoretical frameworks---namely, electromagnetic perturbation theory and a quantise input-output theory---that both accurately predict the behaviour of polarisation-dependent cavity magnon-polaritons. 
In what follows, we start by discussing  electromagnetic perturbation theory as a way to build intuition for the problem.
This theory also allows us to draw direct parallels with well-known concepts such as that of intrinsic handedness of magnetisation precession as well as to calculate the coupling strength without any need for experimental fitting parameters. While intuitive, pertubation theory on its own can only be used to estimate the coupling strength, therefore, by employing a second quantisation model we are also able to obtain reflection (S-parameter) spectra whilst describing the system in the familiar language of magnon-polaritons.
Nonetheless, both models are equivalent and as we will see, describe our experimental data extremely accurately. Moreover, through the polarisation state we are able to engineer an external applied magnetic field non-reciprocity mechanism into cavity-magnon hybridisation, adding yet another tunability mechanism to our system.
Finally, we expect that by being able to fully control and understand the behaviour of these polarisation-dependent systems, we hold considerable promise for realising tunable magnonic devices and enabling advancements in quantum information processing and spin-based technologies.

\section*{Results}
$ $

Our premise is quite simple, spin precession has an inherent chirality; as such, to
effectively excite magnetisation a driving excitation polarised with compatible chirality is needed. 
Therefore, such effect should also manifest when spins in a magnetic sample are driven by fields confined inside a microwave cavity. 

\subsection*{Right and Left Circularly Polarised Cavity Magnon-Polaritons}
$ $

We start by looking at the magnetic response of a material to an applied magnetic field which can be characterised classically by the Landau-Lifshitz (LL) equation \cite{gurevichBook}. 
If we take the case where an external static magnetic field $H_{0}$ is applied along the $z$ direction and a time-varying AC field oscillating the $x$-$y$ plane, represented by $\mathbf{h}e^{i\omega t}$, then the effective field can be written as $\mathbf{H_{eff}} = \hat{\mathbf{z}}H_{0} + \mathbf{h}e^{i\omega t}$ (where the observed magnetic field corresponds to the real part of $\mathbf{H_{eff}}$.)
Upon solving the LL equation, the following relation between the oscillating magnetisation $\mathbf{m}$ and the oscillating magnetic field is derived:

\begin{ceqn}
\begin{align}
\label{mhs}
\begin{bmatrix}
m_x \\
m_y \\
\end{bmatrix} =
\underbrace{\begin{bmatrix}
\chi_{a} & i\chi_{b} \\
-i\chi_{b} & \chi_{a} \\
\end{bmatrix}}_{\overleftrightarrow{\chi}_m(\omega)}
\begin{bmatrix}
h_x \\
h_y \\
\end{bmatrix}.
\end{align}
\end{ceqn}
\noindent Here, $\overleftrightarrow{\chi}_m(\omega)$ is the high-frequency magnetic susceptibility which is a non-symmetric second rank tensor and its components have the form
$\chi_a = \omega_0\omega_m/(\omega_0^2-\omega^2)$ and $\chi_b = \omega\omega_m/(\omega_0^2-\omega^2)$;
where $\omega_0 = \mu_0\gamma H_{0}$ is the natural precession frequency of a magnetic dipole in a constant magnetic field, and $\omega_m = \mu_0\gamma M_s$. 

\begin{figure}
\centering
\includegraphics[width=0.9\linewidth]{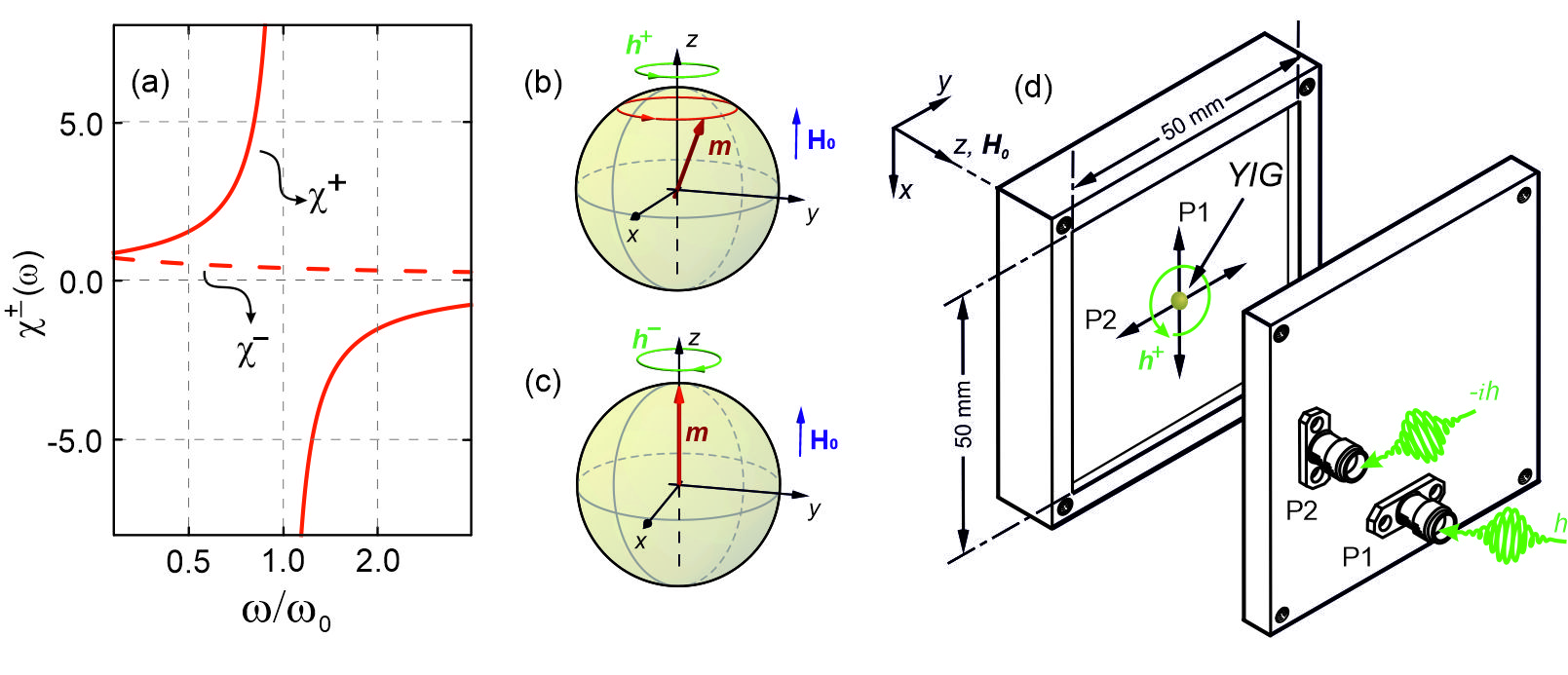}
\caption{\textbf{Compatibility of the handedness of the magnetic driving field with the intrinsic handedness of the magnetisation precession.} (a) Effect of right $\mathbf{h^+}$ and left $\mathbf{h^-}$ circular polarised excitation on susceptibility $\chi^\pm=\omega_m/(\omega_0\mp\omega)$. The lines for $\chi^\pm$ were calculated using the magnetic parameters for YIG: $\mu_0M_s = $0.1758~T, $\gamma = $28~GHz/T, and $H_{0} =$230~mT (corresponding to $\omega_0$ = 6.44~GHz). 
The sphere diagrams in (b) and (c) illustrate the effect of the polarisation of the driving field, $\mathbf{h}$, on the magnetisation, $\mathbf{m}$. (b) Magnetisation excited by a right circularly polarised field inducing a strong response (large precession cone) and (c)  vanishingly-small precession seen for a left circularly polarised excitation. (d) Diagram of the cavity used to generate excitation fields with different polarisations. By controlling the input signals to port 1 (P1) and port 2 (P2), excitation fields with any polarisation can be generated at the sample position (refer to Supplemental Material E for more detail).}
\label{fig:MagPrecess}
\end{figure}

Now, let the driving field be right- or left-circularly polarised, so that it can be written as 
$$\mathbf{h}^+(t) = (\hat{\mathbf{x}}h- \hat{\mathbf{y}}ih)e^{i\omega t}$$
or
$$\mathbf{h}^-(t) = (\hat{\mathbf{x}}h+\hat{\mathbf{y}}ih)e^{i\omega t},$$
respectively. 
Eq.~(\ref{mhs}) can then be rewritten with $\mathbf{m}$ and $\mathbf{h}$ as circularly polarised quantities, and it takes the form: 
\begin{ceqn}
\begin{equation}
    \mathbf{m}^\pm = \frac{\omega_m}{\omega_0 \mp \omega}\mathbf{h}^\pm,
    \label{mcirc}
\end{equation}
\end{ceqn}
where $\mathbf{m}^{\pm}=m_x\pm m_y$ and $\mathbf{h}^{\pm}=h_x\mp ih_y$. 
One of the well-known consequences of this is that for right-circularly polarised driving field $\mathbf{h}^{+}$, the susceptibility $\chi^{+} = \omega_m/(\omega_0 - \omega)$ has a singularity (resonance) at $\omega=\omega_0$ [shown as a solid line in Fig.~\ref{fig:MagPrecess}(a)].
In this case, a large response takes place as the oscillating magnetisation and the excitation vector fields are in phase. 
This is known as the Larmor precession (or Larmor condition) and is sketched in Fig.~\ref{fig:MagPrecess}(b) as a large-angle magnetisation precession cone.
On the other hand, the left-circularly polarised field $\mathbf{h}^{-}$, has a sense of rotation opposite to the magnetisation's natural precession, and thus the Larmor condition cannot be met. Eq.~(\ref{mcirc}) for $\chi^{-}=\omega_m/(\omega_0 + \omega)$ has no singularity and a much smaller magnitude than $\chi^{+}$ [shown as the dashed line in Fig.~\ref{fig:MagPrecess}(a)].
This behaviour is also depicted in Fig.~\ref{fig:MagPrecess}(c), and is known as ``anti-Larmor'' precession \cite{kambersky75,patton03}.
This behaviour is a direct consequence of the handedness of spin precession (i.e. the clockwise or anti-clockwise direction of precession is dictated by the direction of $\mathbf{H_0}$). It should be noted that reversing the direction of $\mathbf{H_0}$ would cause the signs in the denominator of Eq.~(\ref{mcirc}) to flip, changing from $\mp$ to $\pm$. Thus, $\mathbf{h^+}$ would then correspond to a left-circularly polarised excitation with respect to $\mathbf{H_0}$ [Eq.~(\ref{mcirc}) would have no singularity], while $\mathbf{h^-}$ would correspond to right-circularly polarised with respect to $\mathbf{H_0}$ [Eq.~(\ref{mcirc}) would have a singularity at $\omega=\omega_0$].

To ascertain how the behaviour detailed above affects cavity magnon-polariton systems, we have designed a three-dimensional microwave resonator [shown in Fig.~\ref{fig:MagPrecess}(d)] where the polarisation state of the microwave, cavity-bound photons can be controlled using a two-port set up. 
The position of these ports is designed so that---in combination with direct control of the input at each port---the superposition of the excitation vector fields allow for a tunable, all-polarisation state system (i.e. by controlling the input at each port we can can create driving fields with any polarisation state at the sample position). This allows us to experimentally probe the coupling between magnons and cavity photons by placing a Yttrium Iron Garnet (YIG) sphere in the centre of the cavity where the excitation vector fields can be controlled to be left or right circularly polarised by controlling the phase of one of the inputs.

\begin{figure}[ht]
\centering
\includegraphics[width=0.95\linewidth]{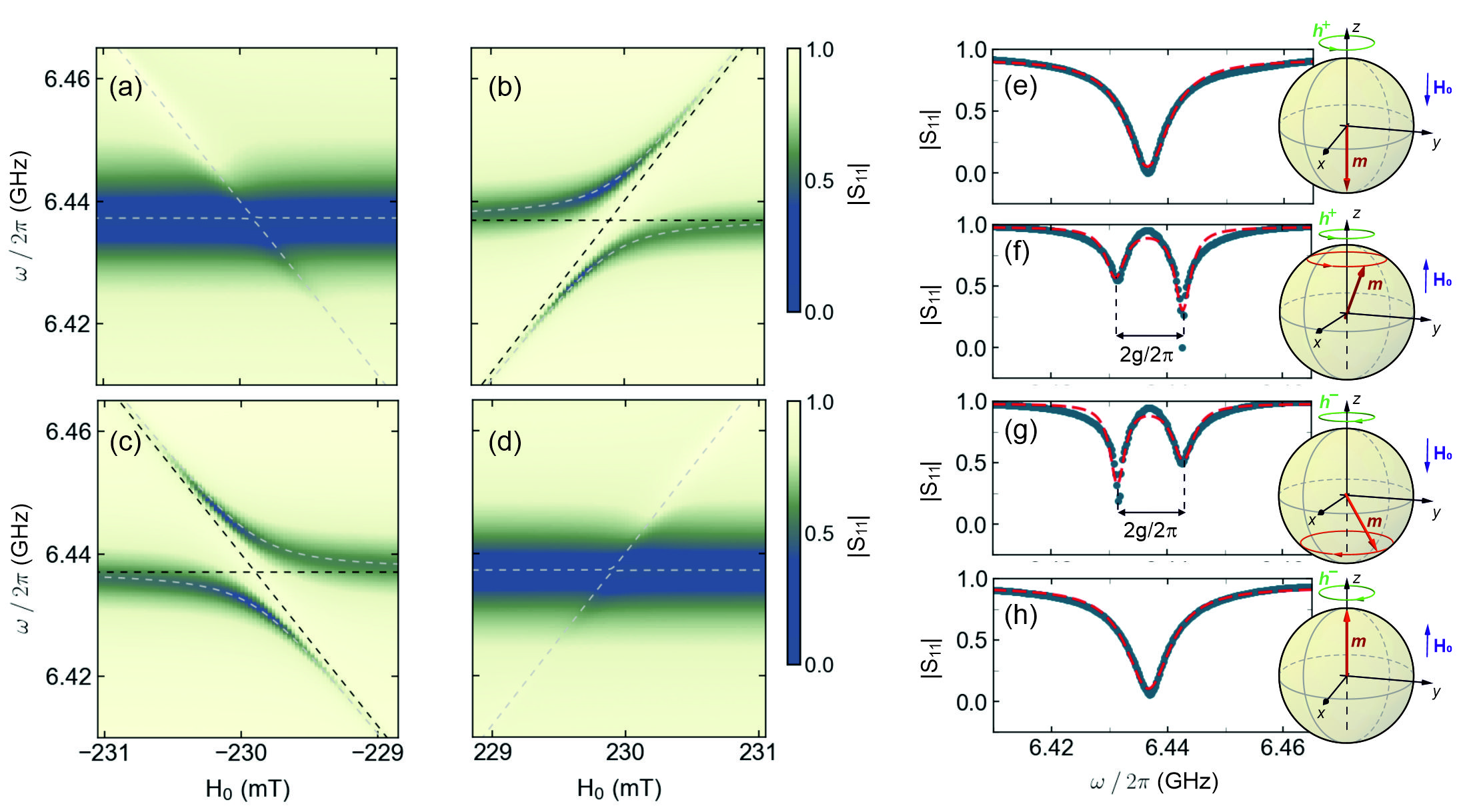}
\caption{
\textbf{Cavity magnon-polariton response to circularly polarised excitation fields and bias fields applied along the $\pm\mathbf{\hat{z}}$ axis}. Normalised experimental spectra and perturbation theory (dashed grey lines) of the quasi-Rabi splitting %~\cite{miller05}
close to $\omega_0 = \omega_c$ for (a) $\mathbf{H_{eff}} = -\hat{\mathbf{z}}H_{0}+\mathbf{h}^+$; (b) $\mathbf{H_{eff}} = \hat{\mathbf{z}}H_{0}+\mathbf{h}^+$; (c) $\mathbf{H_{eff}} = -\hat{\mathbf{z}}H_{0}+\mathbf{h}^-$ and (d) $\mathbf{H_{eff}} = \hat{\mathbf{z}}H_{0}+\mathbf{h}^-$. 
The dashed black lines show the cavity mode and Kittel mode. 
(e)-(h) show the corresponding $|S_{11}|$ parameter measured when $\omega_0 = \omega_c$.
The sphere diagrams illustrate the direction of the magnetic bias field, the excitation field and the subsequent spin precession in the YIG. 
}
\label{fig:CircPol}
\end{figure}

The experimental spectra displayed in Fig.~\ref{fig:CircPol} were obtained by controlling both the excitation field polarisation state and the direction of the bias field, $\mathbf{H_0}$. 
We show examples for left, $\mathbf{h^-}$, and right,  $\mathbf{h^+}$, circularly polarised excitation vector fields for bias fields of $\pm\mathbf{\hat{z}}H_{0}$. Comparing Figs.~\ref{fig:CircPol}(a) and (b), we can see that by simply reversing the direction of the $\mathbf{H_0}$, it is possible to go from level crossing (a) to level repulsion (b).
In the first case, the modes cross each other and no hybridisation takes place while in the second hybridisation takes place between both systems and an 'anti-crossing' of the modes is observed.
In Fig.~\ref{fig:CircPol}(c)-(d) the opposite behaviour is observed; namely, $-\mathbf{H_0}$ shows avoided crossing while $+\mathbf{H_0}$ shows level crossing.
This is achieved by changing the sense of rotation of the circularly polarised excitation vector field with respect to the direction of $\mathbf{H_0}$.

These results can be reproduced using perturbation theory (See Supplemental Material A). 
For this, Maxwell's equations can be used to obtain the following relation to estimate the shift in the cavity resonance frequency $\omega_c$ due to a small perturbation \cite{macedoElectromagneticApproachCavity2021}: 
\begin{ceqn}
\begin{equation}
    \frac{\omega - \omega_c}{\omega_c} = -\frac{\displaystyle\int_{\delta v}\mu_0[\overleftrightarrow{\chi}_m(\omega)\cdot\mathbf{h_c}]\cdot\mathbf{h_c^*}\mathrm{d}v}{2\displaystyle\int_{v}\mu_0|\mathbf{h_c}|^2 \mathrm{d}v},
\label{eq:perturbationTheory}
\end{equation}
\end{ceqn}

\noindent where $\mathbf{h_c}$ denotes the oscillating magnetic field of the cavity, $\delta v$ is the sample volume and $v$ is the volume of the empty cavity. Up to now, perturbation theory has been limited to describing the coupling between magnons and microwave photons that are linearly polarised.
However, if we introduce circularly polarised excitation vector field by making $\mathbf{h_c}=\mathbf{h^\pm}$  at the sample position we can rewrite Eq.~(\ref{eq:perturbationTheory}) as
\begin{ceqn}
\begin{equation}
\label{w-w0_circ}
    \frac{\omega-\omega_c}{\omega_c} =-\frac{\omega_m}{\omega_0\mp \omega } \frac{ \displaystyle\int_{\delta v}|\mathbf{h^\pm}|^2\mathrm{d}v}{2\displaystyle\int_{v}|\mathbf{h_c}|^2 \mathrm{d}v}
\end{equation}
\end{ceqn}
where `$-$' in the denominator is for $\mathbf{h^+}$ and `$+$' is for $\mathbf{h^-}$ (both defined with respect to $\mathbf{H_0}=+\mathbf{\hat{z}}H_0$). Note that we cannot replace $\mathbf{h_c}$ by $\mathbf{h^\pm}$ in the denominator as the integral is evaluated across the whole volume of the cavity and at some positions, unlike at the centre, $\mathbf{h_c}$ may not add up to a circularly polarised quantity. 
These imply that a circularly polarised microwave magnetic field with the same chirality as the precession motion has a singularity at $\omega = \omega_0$; here corresponding to level repulsion. 
On the other hand, the response of the medium to an excitation with opposite chirality to the magnetisation is non-resonant [Eq.~(\ref{w-w0_circ}) and has no singularity if we take the `$+$' sign], which corresponds to coupling annihilation as the solutions of Eq.~(\ref{w-w0_circ}) yields simply the decoupled $\omega_c$ and $\omega_0$ frequencies. 
This is not only in agreement with the experimental results but also directly reflects the behaviour discussed in Fig.~\ref{fig:MagPrecess}.

By solving Eq.~(\ref{w-w0_circ}) for $\omega \approxeq \omega_c, \omega_0$, we obtain 
the eigenfrequencies of the cavity-magnon hybrid system. 
By using the magnetic parameters for YIG [same as used in Fig.~\ref{fig:MagPrecess}(a)] we obtain the eigenfrequencies for the two modes (branches), which are in excellent agreement with the experimental contour data.
In Figs.~\ref{fig:CircPol}(e)-(h), we show cross-sectional $|S_{11}|$ spectra taken from the data in the maps shown in (a)-(d). 
These are taken at a $H_0$ value corresponding to $\omega_c=\omega_0$ since the width of the splitting between the eigenfrequencies at this point is related to the macroscopic coupling strength, $g$ ($2g = \Delta\omega$) \cite{harder16}.
From the plots, we can obtain an experimental value for $\Delta\omega$, and thus the coupling strength, $g$.
Notably, in the case of Fig. \ref{fig:CircPol}(e) and of Fig. \ref{fig:CircPol}(h), the cavity and magnon modes coalesce, i.e., indistinguishable eigenmodes. %, as expected from the driving conditions. 
However, for Fig. \ref{fig:CircPol}(f) and Fig. \ref{fig:CircPol}(g), we measure $\Delta\omega/2\pi$ of 11~MHz experimentally.
Conspicuously, we have a large $\Delta\omega$ for level repulsion, signifying strong coupling, whereas $\Delta\omega=g=0$ for level crossing, indicating coupling annihilation. 
We note that the above analysis has neglected damping of both cavity and magnon modes.
While these can be phenomenologically included in Eq.~(\ref{w-w0_circ})\cite{macedoElectromagneticApproachCavity2021}, since both YIG and the cavity have extremely small dissipation rates, their effect on perturbation theory eigenfrequencies can be neglected.

Furthermore, it is important to note that field non-reciprocity can be induced in our system. 
For instance, in Figs.~\ref{fig:CircPol}(e) and (f), we observe the interaction of a right-circularly polarised excitation, $\mathbf{h}^+$, with spin precession when a static magnetic field, $\mathbf{H_{0}}$, is applied along $\pm\hat{\mathbf{z}}$. 
In Fig. \ref{fig:CircPol}(f), the resonance condition is met as both the wave, $\mathbf{h}^+$, and the magnetisation precession have the same chirality and level splitting is observed (two distinct dips) at $\omega_0=\omega_c$. 
In Fig. \ref{fig:CircPol}(e) on the other hand, the condition is not met as torque now dictates that precession motion has opposite chirality to $\mathbf{h}^+$ and thus only one dip is observed (corresponding to $\omega_c$ as $\omega_0$ is not excited). % (depicted as the orange arrow).
Effectively, by not changing the chirality of the excitation vector field but reversing the sign of $\mathbf{H_{0}}$, it is possible to switch from strong spin precession to almost no precession at all and consequently from strong coupling to coupling annihilation. 
The opposite behaviour then takes place for $\mathbf{h^-}$; namely strong coupling for 
when $\mathbf{H_{0}}$ is applied along $-\hat{\mathbf{z}}$ [Fig.~\ref{fig:CircPol}(g)] and coupling annihilation when $\mathbf{H_{0}}$ is applied along $+\hat{\mathbf{z}}$ [Fig.~\ref{fig:CircPol}(h)].

\subsection*{Tunable magnon-polariton coupling}
$ $

%The subsequent section presents experimental and analytical findings, showcasing the outcomes of changing the amplitude ratio, $\delta$, and phase, $\varphi$.
Thus far, our analysis has primarily focused on the special case of circularly polarised excitation vector fields. 
However, by implementing the experimental set-up shown in Fig.~\ref{fig:MagPrecess}(d), we establish a robust system that enables precise control of the polarisation of the driving excitation vector field.
This allows us to engineer a continuous change in polarisation states (from linearly polarised to circularly and elliptically polarised excitations) by controlling both the relative amplitude and phases between the inputs in both cavity ports so that the oscillating driving field is:
\begin{ceqn}
\begin{equation}
\label{drivingField}
\mathbf{h}_c = (\hat{\mathbf{x}}+ \hat{\mathbf{y}} \delta e^{i \varphi})h e^{i \omega t}
\end{equation}
\end{ceqn}
where $\delta$ is the amplitude ratio and $\varphi$ is the relative phase between the ports (and hence $x$ and $y$ components of $\mathbf{h_c}$ at the sample position) and with observable field corresponding to Re($\mathbf{h_c}$). 
The amplitude ratio $\delta$ and relative phase $\varphi$ are defined as $|h_{cy}|/|h_{cx}|$ and $\arg(h_{cy}) - \arg(h_{cx})$ respectively. 
For context, in this general case, when the amplitude ratio $\delta = 1$ (meaning both components of the excitation field have the same amplitudes) and a phase difference of $\varphi=-90^\circ$, it corresponds to a right-circular polarised excitation ($\mathbf{h^+}$), while a phase difference of $\varphi=+90^\circ$ corresponds to left-circular polarised excitation, ($\mathbf{h^-}$). %in Jones notation
Because we have vast control of the values of $\delta$ and $\phi$, other intermediate polarisation states of the excitation vector fields are allowed to exist, such as elliptically or linearly polarised states. 

This can be readily implemented into Eq.~(\ref{eq:perturbationTheory}) to obtain a general expression similar to Eq.~(\ref{w-w0_circ}) but that takes into account any polarisation state (See Supplemental Material A). However, perturbation theory, by itself, is limited to calculating eigenfrequencies---which can in turn be used to estimate the coupling strength.
Therefore, to quantify and analyse this general case in terms of magnon-polariton modes, we proceed to quantise the cavity fields. 
Unlike electromagnetic perturbation theory, this method relies on more experimental fitting parameters. However, its versatility and completeness allows us not only to calculate the eigenfrequencies of the system but also to directly calculate the reflection (S-parameter) where both magnon damping and cavity losses play a vital role and for the remainder are both included properly. 
The Hamiltonian of the microwave cavity interacting with the magnetic sphere can be written as
\begin{ceqn}
\begin{equation}
\label{eq7.1_main}
   H= \int_v \Big(\mu_0\mathbf{h_c}^2\Big)d^3r-\int_{\delta v}\Big(\mu_0M_sH_0 \Big)d^3r-\int_{\delta v}\mu_0\Big(\mathbf{m}\cdot\mathbf{h_c}\Big)d^3r.  
\end{equation}
\end{ceqn}
The first term in Eq. (\ref{eq7.1_main}) denotes the energy of the cavity of volume $v$ with associated magnetic field $\mathbf{h_c}$ while the second term represents the Zeeman energy. As discussed previously, the static magnetic field $H_0$ is taken to be along $\hat{z}$. The final term in Eq. (\ref{eq7.1_main}) constitutes the cavity-magnon interaction. 
Here, $\mathbf{m}=\gamma\hslash\mathbf{S}/\delta v$ is the magnetization of a YIG sphere of volume $\delta v$ with $S \equiv |\mathbf{S}|=\frac{5}{2}\rho \delta v $ being the coarse-grained, classical spin obtained upon averaging the spin density $\rho$ over the complex unit cell of a material such as YIG \footnote{Note that the magnetic moment of a YIG arises from the Fe$^{3+}$ ions, wherein, two of five iron spins and the remaining three in a unit lattice orient the opposite direction with a net spin $s = 5/2$ \cite{stancil2009spin,li2018magnon} with a spin density that goes as $\rho\approx 4.22 \times 10^{27}$.}. 
The cavity magnetic field $\mathbf{h_c}$ can then be associated with the quantised vector potential of the electromagnetic field in Coulomb gauge \cite{scully1999quantum,agarwal2012quantum,spohn2004dynamics}, which allows to rewrite 
\begin{ceqn}
\begin{equation}\label{eq:hc_q}
    \mathbf{h_c}=i\sqrt{\frac{2\pi\hslash\omega}{v}}\hat{\epsilon}f(\Vec{r})e^{-i\omega t}a+\text{h.c.},
\end{equation}
\end{ceqn}
where $\hat{\epsilon}=\epsilon_x\hat{x}+\epsilon_y\hat{y}$ with $|\epsilon_x|^2+|\epsilon_y|^2=1$, $\omega$ denotes the cavity resonance frequency, $f(\Vec{r})$ describes the spatial mode of the cavity field and $a$ ($a^{\dagger}$) characterizes the annihilation (creation) operators of the cavity field. To describe our experimental setup, in which, the driving magnetic field at the sample position is prepared in a superposition of $\hat{x}$ and $\hat{y}$ with an overall phase shift $\varphi$ and amplitude $\delta$ between $h_{cx}$ and $h_{cy}$, we define the field polarisation as 
\begin{ceqn}
\begin{equation}\label{eq:pol_q}
    \hat{\epsilon}=\frac{\delta e^{i\phi}\hat{x}+\hat{y}}{\sqrt{1+\delta^2}}.
\end{equation}
\end{ceqn}
Employing the Holstein-Primakoff transformation in the dilute magnon limit (since the temperature is far below YIG critical temperature)\cite{holstein1940field}, we can bosonize the spin system by rewriting
\begin{ceqn}
\begin{equation}\label{eq7.2_main}
    S_z=S-b^\dagger b, \quad S_+=S_x+iS_y=\sqrt{2S-b^\dagger b}b \approx \sqrt{2S}b,
\end{equation}
\end{ceqn}
where $b$ $(b^\dagger)$, with  $[b, b^\dagger]=1$, denotes the annihilation (creation) operator of the Kittel mode, i.e., the uniformly precessing magnon mode. Under the rotating wave approximation, using Eqs. (\ref{eq:hc_q}-\ref{eq7.2_main}), we can rewrite Eq. (\ref{eq7.1_main}) as
\begin{ceqn}
\begin{equation}\label{eq7.4_main}
    H=\hslash\omega a^\dagger a+\hslash \omega_0 b^\dagger b+\hslash [\Tilde{g}ba^\dagger+\Tilde{g}^*ab^\dagger].
\end{equation}
\end{ceqn}
For simplicity, $\Tilde{g}$ has been introduced to represent the coefficients multiplying the terms $ba^\dagger$ and $ab^\dagger$, obtained after applying the rotating wave approximation and the substitutions from Eqs. (\ref{eq:hc_q}-\ref{eq7.2_main}) to Eq. (\ref{eq7.1_main}). Thus, $\Tilde{g}$ is the effective magnon-photon coupling given by
\begin{ceqn}
\begin{equation}\label{gtild}
\Tilde{g}=\frac{g}{\sqrt{1+\delta^2}}(1-i\delta e^{i\varphi})
\end{equation}
\end{ceqn}
with $g=\gamma\eta\sqrt{5N\pi \hslash \omega/(2v)}$, $N=\rho \delta v$ and the spatial overlap integral $\eta=\frac{1}{\delta v}\int_{\delta v}f(\Vec{r})d^3r$.
%-------------------------------------------------------------------------------------------------------------
%\section{Input-output theory of the cavity and reflection}\label{sec3}

Finally, the Hamiltonian of the system, incorporating the interaction of the cavity with the travelling waves is given by

\begin{ceqn}
\begin{equation}
    \label{eq15_main}
    H=\hslash\omega a^{\dagger}a+\hslash\omega_0 b^{\dagger}b+\hslash\Tilde{g}a^{\dagger}b+\hslash\Tilde{g}^{*}a b^{\dagger}+i\hslash\epsilon_{c}(a^{\dagger}e^{-i\omega_c t}-ae^{i\omega_c t}),
\end{equation} 
\end{ceqn}
where the last term on the right-hand side of Eq. (\ref{eq15_main}) describes the electromagnetic probe field at frequency $\omega_c$ and power $D_c$, where $\epsilon_c=\sqrt{2\kappa D_c/(\hslash \omega_c)}$ parameterises the strength of the probe field and $\kappa$ depicts the effective decay of the mode $a$ into the electromagnetic vacuum.

Here we use the input-output relation \footnote{Refer to Supplemental Material B}
\begin{ceqn}
\begin{equation}\label{eq14_main}
    a_{in}+a_{out}=\sqrt{2\kappa}a.
\end{equation} 
\end{ceqn}
where $a_{in}$ and $a_{out}$ can be interpreted as the incoming excitation into the cavity and outgoing signal, respectively. Therefore, $a_{in}=\varepsilon_c/\sqrt{\kappa}$.

The dynamics of the system in the frame rotating at frequency $\omega_c$ is governed by the following Heisenberg equations (written in a matrix form):

\begin{ceqn}
\begin{align}
\label{eq:11_main}
\begin{bmatrix}
\dot{a} \\
\dot{b} \\
\end{bmatrix} =
-i\begin{bmatrix}
\omega_c-\omega-i\kappa & \Tilde{g} \\
\Tilde{g}^{*} & \omega_0-\omega-i\eta \\
\end{bmatrix}
\begin{bmatrix}
a \\
b \\
\end{bmatrix}
+
\begin{bmatrix}
\varepsilon_c \\
0 \\
\end{bmatrix}.
\end{align}
\end{ceqn}
where $\eta$ is the decay rate of the Kittel mode of the YIG sphere.

In the long-time limit, we can solve the Eq.~(\ref{eq:11_main}) to obtain $a$. Substituting this into Eq.~(\ref{eq14_main}), we find the reflection coefficient to be
\begin{ceqn}
\begin{equation} \label{S_11}
    |S_{11}(\omega)|=\Big|\frac{a_{out}}{a_{in}}\Big|=\Big|{\frac{i\kappa(\omega_0-\omega-i\eta)}{(\omega-\omega_c+i\kappa)(\omega_0-\omega-i\eta)+|\Tilde{g}|^2}-1}\Big|.
\end{equation}
\end{ceqn}
It is worth mentioning that the $\varphi$ dependence of $S_{11}$ from the $|\Tilde{g}|^2$ in the denominator is identical to the $\varphi$ dependence in Eq. (\ref{perturbation_w_WpWC}), ensuing from the full perturbation theory discussed in the supplementary material, further corroborating the equivalence of both methods.

\begin{figure}[ht]
\centering
\includegraphics[width=0.96\linewidth]{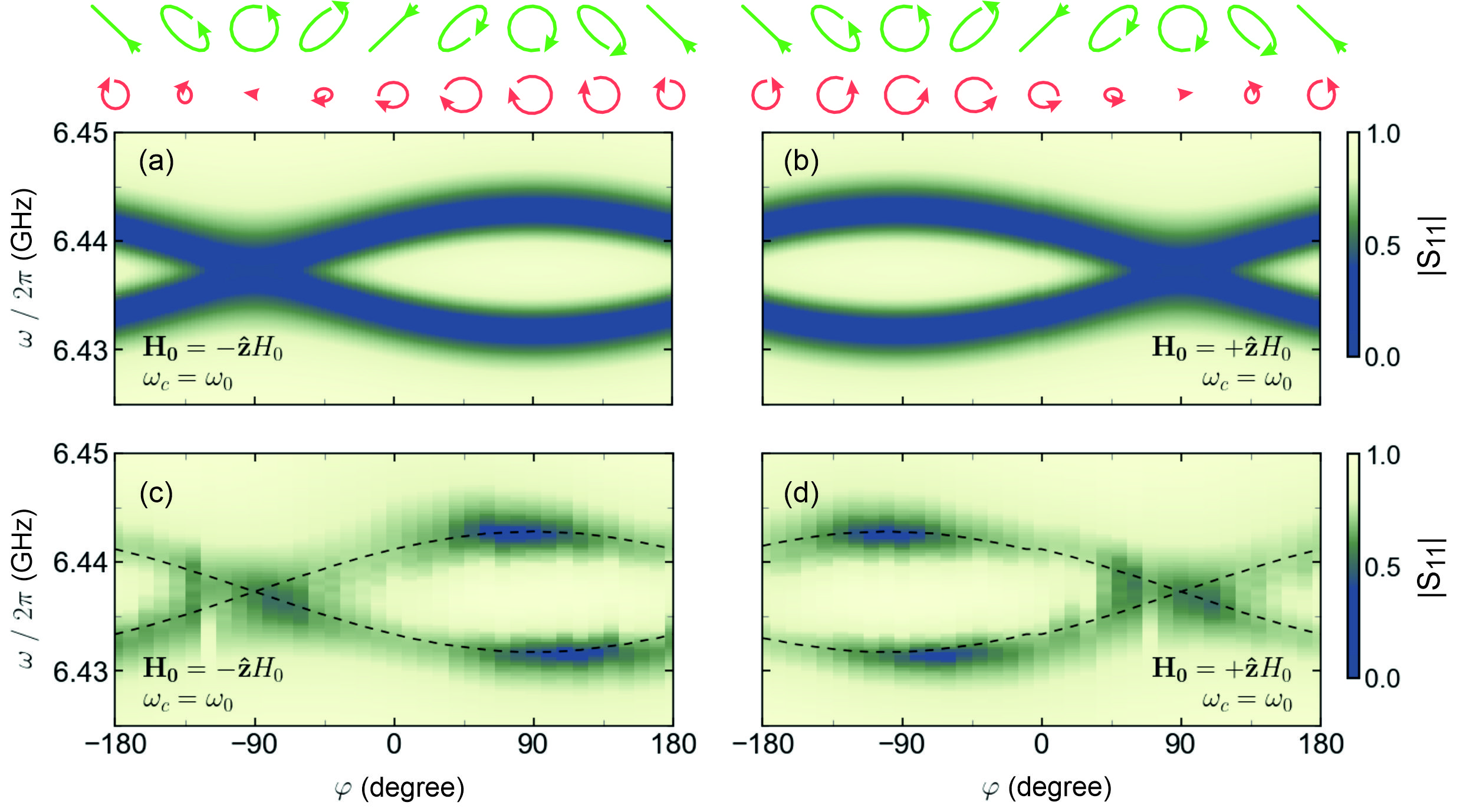}
\caption{\textbf{Tunable hybridisation of cavity magnon-polaritons.} The hybridisation behaviour at $\omega_0 = \omega_c$ for a bias field of $+\hat{\mathbf{z}}H_{0}$ in (a) and $-\hat{\mathbf{z}}H_{0}$ in (b) as $\varphi$ is changed and $\delta=1$. The values, scaled between 0 and 1, were calculated from Eq. (\ref{S_11}) using the experimentally measured parameters $\kappa/2\pi = 4.45$~MHz, $\eta/2\pi = 0.7$~MHz (see Supplemental Material D) and $g/2\pi = 3.9$~MHz. The experimental amplitudes of the normalised $|S_{11}|$ parameter for various 
values of $\varphi$ when $\omega_0 = \omega_c$ and $\delta=1$ is shown for a bias field of $+\hat{\mathbf{z}}H_{0}$ in (c) and $-\hat{\mathbf{z}}H_{0}$ in (d). %The dashed black lines are calculated using perturbation theory [the solutions from Eq.~(\ref{ellipticalS11})]. 
The dashed black lines are the eigenfrequencies extracted from (a) and (b).
The diagram above the figure illustrates the excitation field polarisation (depicted as the green arrows) and the subsequent precession cone (depicted as the red arrows) as $\varphi$ is changed, for $\delta =1$, calculated from the LL equation.
}\label{fig:ElliptPol}
\end{figure}

From Eq.~(\ref{S_11}), and from the expression for $\Tilde{g}$, one can expect that by manipulating the polarisation of the cavity excitation vector fields, the cavity-magnon interaction can be controlled.
To demonstrate this, in Fig.~\ref{fig:ElliptPol}, we show a map of $|S_{11}(\omega)|$, comparing spectra at $\omega_0=\omega_c$ and $\delta=1$ for a $360^\circ$ phase scan, illustrating, how to tune the Rabi splitting by simply varying $\varphi$, even when the amplitude of both $h_{cx}$ and $h_{cy}$ are identical. 
In Figs.~\ref{fig:ElliptPol}(a) and (b) we show calculated values of transmission coefficient using Eq.~(\ref{S_11}) which are directly compared with experimental data given in Figs.~\ref{fig:ElliptPol}(c) and (d).
The theory shows remarkable agreement with the experimental contours and from both cases, it is evident that the Rabi splitting increases as we move from $\varphi = 0$ to $-90^\circ$ and decreases towards the annihilation point at $\varphi = +90^\circ$ for $+\mathbf{\hat{z}}H_0$ [Figs.~\ref{fig:ElliptPol}(b) and (d)].
These points correspond to $\mathbf{h^+}$ and $\mathbf{h^-}$, respectively and are defined with respect to the direction of $H_0$---the cases discussed in Fig.~\ref{fig:MagPrecess}.
The diagram above the figure depicts the polarisation of the excitation state for various $\varphi$ (green) and the subsequent magnetic precession in the sample (red). 
For a bias field of $\mathbf{H_0}= +\hat{\mathbf{z}}H_{0}$, we observe a significant decrease in the precession cone at $+90^\circ$, consistent with previous findings indicating no coupling due to the opposite chirality of the excitation field relative to the magnetisation precession handedness.
As the phase is swept to $-90^\circ$, the chirality of the magnetic excitation field progressively matches the intrinsic handedness of the magnetisation precession. At $\varphi = -90^\circ$, both the right-circularly polarised excitation field and magnetisation have the same chirality, enabling maximal driving of the precession dynamics.
In addition to the cases of circularly polarised excitation, other special cases include $\varphi=-180^\circ, 0^\circ, 180^\circ$ which result in a linearly polarised $\mathbf{h_c}$.
The result for these cases is somewhat similar to right circular polarisation, however, the precession cone is smaller for linear polarisation due to the weaker coupling between the excitation field % because the excitation field does not couple as strongly 
and magnetisation.  
For linearly polarised excitation, we obtained a $\Delta\omega/2\pi=7.8$~MHz (which is a factor of $1/\sqrt{2}$ smaller when compared with the $\Delta\omega$ for $\mathbf{h^+}$).
Moreover, any other case then represents elliptical polarisation; where the behaviour is analogous to that of circularly polarised states (in that the handedness is important) albeit the coupling is not as strong.

Interestingly, switching the magnetic bias field effectively [$-\mathbf{\hat{z}}H_0$, shown in Figs.~\ref{fig:ElliptPol}(b) and (d)] induces a $180^\circ$ phase shift in the wave's rotation, as anticipated. 
Namely, magnetisation is now excited at $\varphi=+90^\circ$ which corresponds to a right circularly polarised excitation with respect to the direction of $H_0$. 
Similarly, the response of the magnetisation greatly decreases, and vanishes, as $\varphi$ approaches $-90^\circ$. 
This is because the excitation field is left circularly polarised in relation to the 
magnetisation and no component of $\mathbf{h_c}$ is able to excite spin precession.

So far, we've been examining the impact of elliptically polarised excitation fields with an amplitude ratio of $\delta=1$. However, our versatile experimental set-up [see Fig.~\ref{fig:MagPrecess}(d)] allows us to explore how the system responds to various excitation fields with all possible polarisation states.
In turn, we can fully understand and control the hybridisation behaviour with excitation fields of any polarisation state.
To this end, we investigate the effect of varying $\delta$ and $\varphi$ to summarise the behaviour of $\Delta\omega$ for all different excitation conditions. It is important to note that in this experiment, we vary the amplitude ratio within the range $0\leq\delta\leq1$. This is because $\delta$ ranging from 1 to infinity corresponds to driving fields with the same polarisation state as $\delta$ spanning from 1 to 0. Thus, in this parameter space, we expect the same hybridisation behaviour (more details are provided in Supplemental Material F).

\begin{figure}[ht]
\centering
\includegraphics[width=0.96\linewidth]{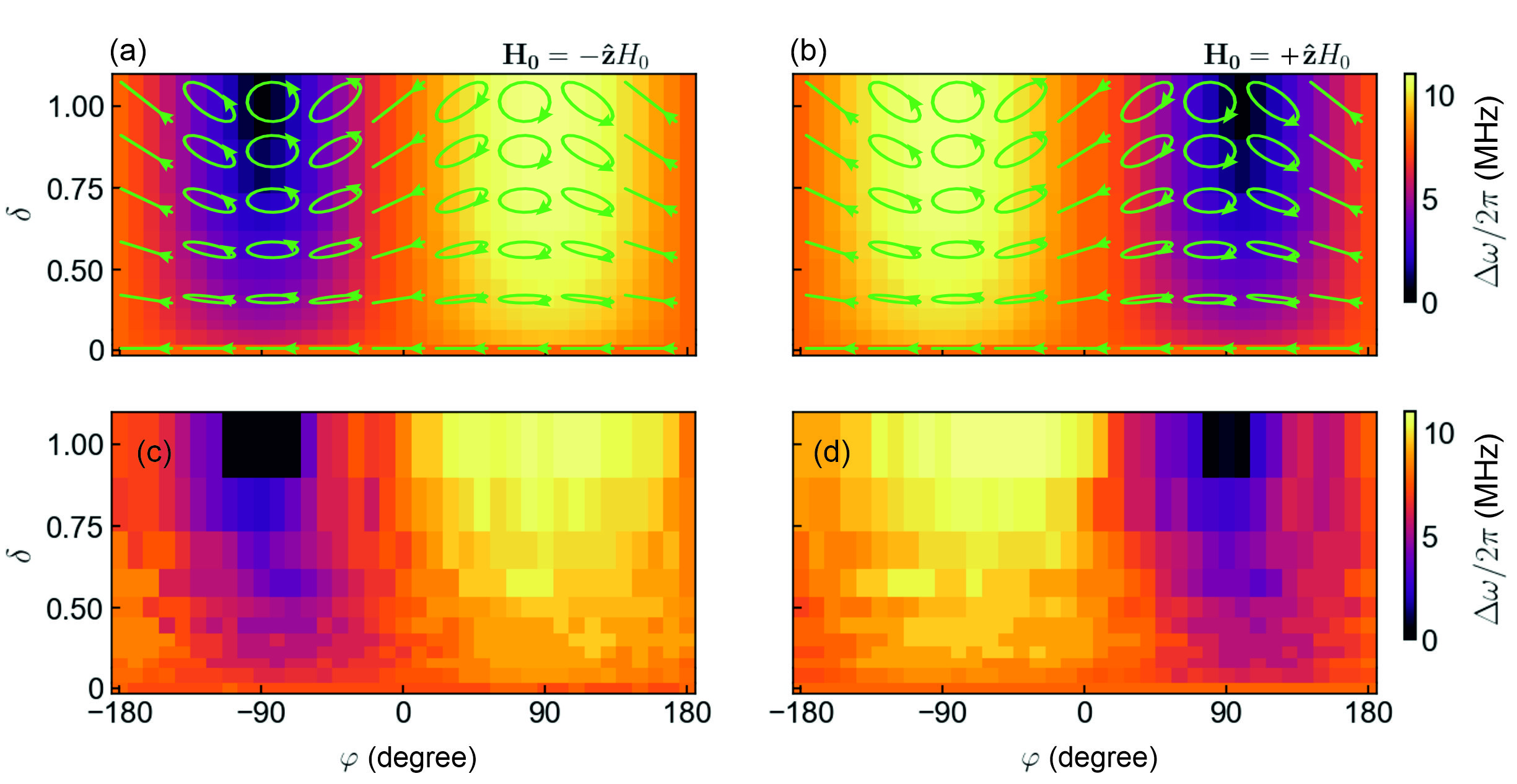}
\caption{\textbf{Summary of the dependence of cavity magnon-polariton coupling behaviour on excitation field polarisations.} 
A heat map summarising the width of the quasi-Rabi splitting, $\Delta\omega/2\pi$, as a function of $\delta$ and $\varphi$ calculated from Eq.~(\ref{S_11}) for a bias field of $+\hat{\mathbf{z}}H_{0}$ in (a) and $-\hat{\mathbf{z}}H_{0}$ in (b). Shown in green, are the polarisation states corresponding to various $\delta$ and $\varphi$. The experimentally measured quasi-Rabi splitting, $\Delta\omega/2\pi$, for various $\delta$ and $\varphi$ for the same bias fields are shown in (c) and (d) respectively.
}
\label{fig:AmpPhase}
\end{figure}

From Fig. \ref{fig:AmpPhase}, we observe the potential to transition between different coupling regimes solely by adjusting the relative phase and amplitude between the two field components -- where any arbitrary value of $\delta$ and $\varphi$ will correspond to an elliptically polarised excitation. The arrows in Fig.~\ref{fig:AmpPhase}(a) and (b) illustrate the polarisation of the excitation field for various $\delta$ and $\varphi$ and the heat maps present the theoretical values of $\Delta\omega/2\pi$ obtained from Eq. (\ref{gtild}) for the corresponding excitation fields, providing a comprehensive mapping of the coupling behaviour across the parameter space of $\delta$ and $\varphi$. 
The experimental results investigating the amplitude ratio and phase effects are presented in Fig. \ref{fig:AmpPhase}(c) and (d). As the amplitude ratio $\delta$ approaches 0, the excitation field becomes increasingly linearly polarised, as evident from the arrow diagrams. Notably, at $\delta = 0$, we measured a $\Delta\omega/2\pi$ of 7.8~MHz, matching the expected width for a linearly polarised excitation.
The analytical prediction of $\Delta\omega$ obtained from Eq.~(\ref{gtild}) demonstrates excellent agreement with our experimental findings as we calculate $\Delta\omega/2\pi = 7.8$~MHz for $\delta = 0$.

We can also see for the special case when the excitation field is right circularly polarised with respect to $\mathbf{H_0}$, the coupling reaches its maximum value. Conversely, when the excitation field is left circularly polarised relative to $\mathbf{H_0}$, we clearly observe no coupling. These findings are in excellent agreement with the results discussed in Fig. \ref{fig:CircPol} and Fig. \ref{fig:ElliptPol}.

\section*{Conclusion}
$ $

Here, we have shown how the excitation vector field can dramatically alter magnon-cavity hybridised states, i.e. cavity magnon-polaritons. 
By all-polarisation control over the excitation vector field, it is possible to not only enhance the coupling but also annihilate it depending on the sense of rotation of the excitation.

The understanding presented here is particularly relevant for technological applications based on cavity magnonics, where controlling the coupling between magnons and photons is crucial. For instance, they are expected to aid
bidirectional conversion between radio-frequency waves and light \cite{hisatomi16,lambert19}. 
Moreover, as cavity magnon–polaritons can also couple with qubits \cite{lachance-quirionEntanglementbasedSingleshotDetection2020}, controlling the interaction between magnons and photons enables the ability to coherently exchange of information between qubits and cavity magnon-polaritons, therby providing a valuable path for quantum information processing \cite{tabuchiCoherentCouplingFerromagnetic2014,lachance_quirion19}. 
In both cases, engineering as well as understanding the coupling are crucial steps to optimise the conversion and (or) information exchange. In our work, we engineered a two-port cavity system where the nature of coupling can be tuned at will. 
Through the theoretical framework presented here, we also are able to predict as well as gain further insight into the nature of the coupling between microwave cavities and magnons in a rigorous manner for any excitation field. 
Recent work has also used two port cavities such as what we do here beyond controlling strong coupling and to obtain another regime of coupling altogether: level attraction \cite{gardin2024}.
In our case, we have obtained spectra resembling level attraction (See Supplementary Note F) due to interference between transmission and reflection specta, which is a limitation of how this set up can generate polarisation. 
However, as shown in Ref. (\citenum{gardin2024}) this versatile experiment can be used well beyond polarisation investigation.
%resonator with any field configuration and for any geometry of the sample (e.g. spheres or thin films, pillars, etc). 

It is also important to point out here that field non-reciprocity can be induced in our system.  
A qualitatively similar behaviour to the non-reciprocity for circularly polarised waves has recently drawn the attention of the optics and photonics community in the context of cavity quantum electrodynamics. In this case, however, the non-reciprocity is obtained through time-reversal symmetry break in Fabry-P\'erot cavities \cite{yang19}. 
While in our case we do not have time-reversal symmetry breaking---as reversal of time would change both the direction of $\mathbf{H_{0}}$ and sense of rotation of $\mathbf{h_c}$---the field reversal behaviour is still of particular interest for non-reciprocal devices as it can enable tunable devices that can readily switch between regimes of hybridisation. For instance, in devices where information is carried from photons to magnons, this field non-reciprocity could be used (i.e. by controlling the direction of the bias field) to selectively enable or block the information exchange.

Finally, we have introduced the excitation vector fields as a way to efficiently sweep through various regimes of hybridisation. In particular, we have shown how the state of polarisation can dramatically affect spin precession inside the cavity resonator. This is particularly relevant, not only from a fundamental point of view but also practically, as understanding the role of the cavity excitation fields in governing the coupling and, consequently, the formation of cavity magnon-polaritons is crucial for effectively engineering and optimising cavity magnonic devices.

\section*{Methods}

\subsection*{Experimental Setup Details}
$ $

In order to obtain more complex field configurations within a microwave cavity resonator, we have 
employed a cavity shown in Fig.~\ref{fig:MagPrecess}(d) featuring equal dimensions along $x$ and $y$ measuring 50$\times$50$\times$5~mm.
A bias field, $H_{0}$ along the $\pm z$ axis, was applied by placing the cavity with the sample between an electromagnet. 
The TE$_{120}$ mode was obtained by exciting port 1 of the cavity, which has an anti-node at the centre.
At this central point, the y-component of the magnetic field, $h_{cy}$, is zero (or much smaller than the x-component). Thus, we can neglect $h_{cy}$, and consider the light to be linearly polarised in the x-direction. 
Placing a small magnetic sample (YIG sphere of diameter 0.25~mm) at the centre of the cavity (at the anti-node of $\mathbf{h_c}$) enabled excitation of spin dynamics with a linearly polarised field. 
By subjecting the setup to an increasing magnetic field, the magnetic resonance frequency of the sample $\omega_0$ also increased until it closely matched the cavity resonance frequency $\omega_c$. In this frequency region, the magnons within the magnetic sample interact with the photons confined within the cavity, creating a connection between the two modes, resulting in observed Rabi splitting.

In order to induce other states of polarisation, we have introduced an additional coupler, labeled as port 2, into the cavity resonator. This second coupler generates an identical mode as the first coupler; however, its position is chosen such that the mode it generates is rotated by 90$^\circ$ relative to the first coupler. Therefore, the cavity field is no longer linearly polarised in the x-direction at the centre, as the second coupler now generates a y-component in the centre.
The superposition of these fields still generates a linearly polarised excitation though. 
However, the direction of the overall oscillating magnetic field in the centre is now rotated by 45$^\circ$. 
By controlling the phase and amplitude of the signal into the second coupler relative to the signal at the first coupler, the y-component of the driving field can be modified, allowing for more complex driving fields, such as circularly polarised light.
To probe this behaviour we place a phase shifter in the path of the second microwave signal which 
can change the phase of the microwaves, $\varphi$, from that port with respect to the other.
We also added a digital attenuator to the path of the signal; this will allow us to change 
the amplitude, $\delta$, of the second port with respect to the first.

\begin{SCfigure}[][ht]
%\centering
\includegraphics[width=0.5\linewidth]{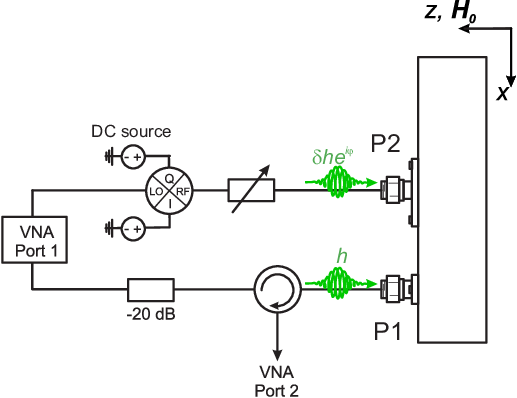}
\caption{\textbf{Circuit diagram for the measurements of the square cavity setup shown in Fig.~\ref{fig:MagPrecess} of the main text.} 
The exciting signal from the VNA is split along two paths. Along the top path, the signal is phase shifted by $\varphi$, and attenuated by $\delta$ before being fed into port 2 on the cavity. 
A fixed attenuator was added along the bottom path, to allow for equal excitation of the two modes in the cavity, before being fed into port 1 of the cavity. Using a circulator on the bottom path, reflection from port 1 (or transmission from port 1 to port 2) is measured on a second port of the VNA.}
\label{fig:ExpGeom}
\end{SCfigure}

Microwave signals were supplied by port 1 of a Rohde \& Schwarz ZVA 40 vector network analyser (VNA) -- a diagram of the setup can be seen in Fig.~\ref{fig:ExpGeom}. 
The signals from the VNA were split along two paths and fed into the two ports of the square cavity [see Fig.~\ref{fig:MagPrecess}(d)]. A static attenuator was added to one of the signal paths to ensure the initial amplitude of the two excitation signals were approximately equal. Along the signal pathway, a circulator was also added. The circulator sent signal excitation to the cavity, but signals reflected from or transmitted through the cavity were sent to port 2 of the VNA. 
To control the relative phase between the two cavity ports, required for polarised excitation, a Marki IQ mixer (MLIQ0416L) was added to the other signal path. The IQ mixer was computer-controlled using a custom voltage source with built-in digital to analogue conversion. %USB-3101 microcontroller
For the frequencies of interest, a map of signal phase and amplitude dependence on DC voltage was generated (See Supplemental Material D). This map was used to calibrate the IQ mixer output. To tune the relative power excitation applied to the cavity ports, an LDA-203B Digital Attenuator was used.

In the experiment, the tones are coupled even in the absence of the YIG sample; there is a finite cross-talk. 
This cross-talk together with the cross-coupling is always present and leads to small deviations from the theory. However, these effects are negligible for the sample size used here and do not affect the understanding of the observed phenomena and general applicability of the theory. More details on the effect of the sample size are given in the supplemental material.

\bibliography{sample}

\section*{Acknowledgements}
This work was supported by the Engineering and Physical Sciences
Research Council (EPSRC) under the project EP/X017850/1, the European Research Council (ERC) under the Grant Agreement 648011, the Initiative and Networking Fund of the Helmholtz Association, the Leverhulme Trust and the University of Glasgow through LKAS funds. 
Work by D. A. Bozhko was supported by the U.S. Department of Energy (DOE), Office of Science, Basic Energy Sciences (BES) under Award \#DE-SC0024400. J. M. P. Nair acknowledges support from DOE under Grant. No DE- SC0024090. B. Flebus acknowledges support from the National Science Foundation under Grant No. NSF DMR-2144086.
M. A. Smith acknowledges funding from EPSRC under grant number EP/S023321/1.
R. Holland was supported by the Engineering and Physical Sciences Research Council (EPSRC) through the Vacation Internships Scheme.
T.Wolz. acknowledges financial support by Helmholtz International Research School for Teratronics (HIRST).
R. M. gratefully acknowledges insightful discussion with Robert E. Camley, Robert L. Stamps, and Karen L. Livesey.

%\noindent For the purpose of open access, the authors have applied a Creative Commons Attribution
%(CC BY) licence to any Author Accepted Manuscript version arising from this submission.

\section*{Author Contributions}
RM, DAB and MPW conceived the experiment.
DAB and LJM performed initial measurements with input from IB and TW.
AJ optimised the experimental set-up with theoretical input from BF and RM and performed the measurements with support from MAS and cavity designs from RH. 
AJ carried out data analysis with contributions from MAS and JMPN. 
JMPN derived the second quantised model and AJ expanded the electromagnetic perturbation theory. 
AJ wrote the manuscript with input from and discussions with all coauthors. 
DAB, BF, MPW, and RM supervised the project.

\section*{Data availability}

All experimental data presented in this study is available in the Enlighten database at
(DOI to be generated by the University of Glasgow once final article is accepted).
\newpage
\clearpage
\renewcommand{\theequation}{S\arabic{equation}}
\renewcommand{\thetable}{S\arabic{table}}
\renewcommand{\thefigure}{S\arabic{figure}}
\setcounter{equation}{0}
\setcounter{table}{0}
\setcounter{figure}{0}
\setcounter{page}{0}
\maketitle

\noindent{\huge\textbf{Supplementary Information for:\vspace{0.3cm}\\The role of excitation vector fields and\vspace{0.3cm}\\ 
all-polarisation state control of cavity magnonics}}
\vspace{0.5cm}

\noindent \textbf{\large Alban Joseph$^1$, Jayakrishnan M. P. Nair$^2$, Mawgan A. Smith$^1$, Rory Holland$^1$, Luke J.
McLellan$^1$, Isabella Boventer$^3$, Tim Wolz$^4$, Dmytro A. Bozhko$^{5}$, Benedetta Flebus$^2$, Martin
P. Weides$^1$, and Rair Mac\^edo$^{1,*}$}
\vspace{0.5cm}

\noindent $^1$James Watt School of Engineering, Electronics \& Nanoscale Engineering Division, University of Glasgow, Glasgow
G12 8QQ, United Kingdom

\noindent $^2$Department of Physics,Boston College,140 Commonwealth Avenue,Chestnut Hill, MA 02467
\noindent $^3$Unite Mixte de Physique CNRS, Thales, University Paris-Sud, Universite Paris-Saclay, F-91767 Palaiseau, France

\noindent $^4$Institute of Physics, Karlsruhe Institute of Technology, 76131 Karlsruhe, Germany

\noindent $^5$Center for Magnetism and Magnetic Materials, Department of Physics and Energy Science, University of Colorado
Colorado Springs, Colorado Springs, Colorado 80918, USA

\noindent $^*$Rair.Macedo@glasgow.ac.uk
\thispagestyle{empty}
\newpage

\section*{Supplementary Note A: Perturbation Theory}
$ $

Maxwell's equations can be used to obtain the following relation to estimate the shift in the cavity resonance frequency $\omega_c$ due to a small perturbation\cite{macedoElectromagneticApproachCavity2021}:
\begin{ceqn}
\begin{equation}
    \frac{\omega - \omega_c}{\omega_c} = -\frac{\displaystyle\int_{\delta v}\mu_0[\overleftrightarrow{\chi}_m(\omega)\cdot\mathbf{h_c}]\cdot\mathbf{h_c^*}\mathrm{d}v}{2\displaystyle\int_{v}\mu_0|\mathbf{h_c}|^2 \mathrm{d}v}
\label{eq:perturbationTheorySupplement}
\end{equation}
\end{ceqn}
At the sample position, if we write the a cavity excitation vector field in its most general form of $\mathbf{h_c} = (\hat{\mathbf{x}}+ \hat{\mathbf{y}} \delta e^{i \varphi})h$ Eq.~(\ref{eq:perturbationTheorySupplement}) can be also generalised and rewritten as:

$$\frac{\omega - \omega_c}{\omega_c} =-\frac{ \mu_0\displaystyle\int_{\delta v}[\chi_{a} |h|^2 + \chi_{a} \delta^2 |h|^2 + i\chi_{b} (h^*\delta he^{i \varphi} - h \delta h^*e^{-i \varphi} )]\mathrm{d}v}{2\displaystyle\int_{v}\mu_0|\mathbf{h_c}|^2 \mathrm{d}v}$$

\noindent or simply:

\begin{ceqn}
\begin{equation}
\label{eq:pertBeforeRCP}
\frac{\omega - \omega_c}{\omega_c} = -\Big[\chi_{a}(1+\delta^2)-2\chi_{b}\delta\sin(\varphi)\Big] \frac{ \displaystyle\int_{\delta v}\mu_0 |h|^2\mathrm{d}v}{2\displaystyle\int_{v}\mu_0|\mathbf{h_c}|^2 \mathrm{d}v}.
\end{equation}
\end{ceqn}

\noindent For the simple case of a  ferromagnet, we can substitute the components of $\overleftrightarrow{\chi}_m(\omega)$ into Eq. (\ref{eq:pertBeforeRCP}) to obtain:

$$
\frac{\omega - \omega_c}{\omega_c} = -\frac{\omega_0 \omega_m}{\omega_0^2 -\omega^2}\Big[(1+\delta^2)-2\frac{\omega}{\omega_0}\delta \sin(\varphi) \Big] \frac{ \displaystyle\int_{\delta v}\mu_0 |h|^2\mathrm{d}v}{2\displaystyle\int_{v}\mu_0|\mathbf{h_c}|^2 \mathrm{d}v}$$

\noindent In the limit close to $\omega = \omega_0$ we obtain:
\begin{ceqn}
\begin{equation*}
\frac{\omega - \omega_c}{\omega_c} = -\frac{\omega_0 \omega_m}{\omega_0^2 -\omega^2}\Big[ 1+\delta^2-2\delta \sin(\varphi)\Big] \frac{ \displaystyle\int_{\delta v}\mu_0 |h|^2\mathrm{d}v}{2\displaystyle\int_{v}\mu_0|\mathbf{h_c}|^2 \mathrm{d}v}
\end{equation*}
\end{ceqn}
or simply:
\begin{ceqn}
\begin{equation}
\label{perturbation_w_WpWC}
    \frac{\omega - \omega_c}{\omega_c} = - \frac{\omega_0 \omega_m}{\omega_0^2-\omega^2}\frac{W_p}{W_c},
\end{equation}
\end{ceqn}

Where $W_p = [1 + \delta^2 + 2 \delta \sin\varphi] \int_{\delta v}  \mu_0|h_{c}|^2 \mathrm{d}v$ is the magnetic energy at the sample position and $W_c = 2 \int_{v} \mu_0|\mathbf{h_c}|^2 \mathrm{d}v$ defines the total cavity energy, both measured in Joules. 
These quantities do not require experimental input and can be estimated using numeric solvers (such as COMSOL multiphysics) for any type of resonators, including 2D integrated devices. \cite{macedoElectromagneticApproachCavity2021} 
Here, however, as we have a simple, rectangular resonator, an analytical expression for $W_c$ can also be derived as follows:

$$W_c = 2 \int_{v} \mu_0|\mathbf{h_c}|^2 \mathrm{d}v$$

$$W_c = 2 \mu_0 \int_{0}^{a} \int_{0}^{b} \int_{0}^{c} \mathbf{h_c^*} \cdot \mathbf{h_c} \, \mathrm{d}x \, \mathrm{d}y \, \mathrm{d}z$$
where $a$, $b$ and $c$ are the x, y and z dimensions of the cavity, respectively. Since the cavity mode is a superposition of the TE$_{120}$ and TE$_{210}$ mode, the oscillating magnetic field in the cavity can be written as

$$
\mathbf{h_c} =
\begin{bmatrix}
h_{x120}  \\
h_{y120} \\
%h_{z120} \\
\end{bmatrix}
+
\begin{bmatrix}
h_{x210} \\
h_{y210} \\
%h_{z210} \\
\end{bmatrix} \delta e^{i \varphi}
$$

\noindent We only consider the $x$ and $y$ components here since no magnetic field component exists in the $z$ direction for both these modes. 

Therefore,

$$\mathbf{h_c^*} \cdot \mathbf{h_c} = |h_{x120}|^2 + h_{x120}h_{x210}^*\delta e^{-i \varphi} + h_{x210}h_{x120}^*\delta e^{i \varphi} + \delta^2|h_{x210}|^2 + |h_{y120}|^2 + h_{y120}h_{y210}^*\delta e^{-i \varphi} + h_{y210}h_{y120}^*\delta e^{i \varphi} + \delta^2|h_{y210}|^2 %+ |h_{z120}|^2 + h_{z120}h_{z210}^*\delta e^{-i \varphi} + h_{z210}h_{z120}^*\delta e^{i \varphi} + \delta^2|h_{z210}|^2
$$

\noindent The magnetic field distribution for a TE$_{mnl}$ mode in a rectangular cavity can be written as

$$h_{cx} = iA\frac{\kappa_{0y}}{\omega_c\mu_0}\sin{\kappa_{0x}x\cos{\kappa_{0y}y}}$$
$$h_{cy} = -iA\frac{\kappa_{0x}}{\omega_c\mu_0}\cos{\kappa_{0x}x\sin{\kappa_{0y}y}}$$
%$$h_{cz} = A\cos{\kappa_{0x}x\cos{\kappa_{0y}y}}$$
where $\kappa_{0x} = m \pi / a, \kappa_{0y} = n \pi / b$ and A is an amplitude constant.

With these definitions:

$$\int_{0}^{a} \int_{0}^{b}  h_{x210}h_{x120}^* \, \mathrm{d}x \, \mathrm{d}y = \int_{0}^{a} \int_{0}^{b}  h_{x120}h_{x210}^* \, \mathrm{d}x \, \mathrm{d}y  = \int_{0}^{a} \int_{0}^{b}  h_{y210}h_{y120}^* \, \mathrm{d}x \, \mathrm{d}y  = \int_{0}^{a} \int_{0}^{b}  h_{y120}h_{y210}^* \, \mathrm{d}x \, \mathrm{d}y  = 0$$

\noindent Therefore $W_c$ can be written as

$$W_c = 2 \mu_0 \int_{0}^{a} \int_{0}^{b} \int_{0}^{c} |h_{x120}|^2 + \delta^2|h_{x210}|^2 + |h_{y120}|^2 + \delta^2|h_{y210}|^2 \, \mathrm{d}x \, \mathrm{d}y \, \mathrm{d}z$$

\noindent By substituting field distributions we get

$$W_c = 2 \mu_0 \int_{0}^{c} \frac{1}{\mu_0^2 \omega_c^2} \frac{5 \pi^2 (1 + \delta^2)}{4}\, \mathrm{d}z$$

$$W_c = \frac{5 \pi^2 \left(1 + \delta^2 \right)c}{2 \mu_0 \omega_c^2}$$

We can solve Eq.~(\ref{perturbation_w_WpWC}) for $\omega$ close to both $\omega_c$ and $\omega_0$, and obtain an analytical expression for the eigenfrequencies. Since we are able to include the new terms due to $\delta$ and $\varphi$ into $W_p$ and $W_c$, our resulting equation remain the same as that obtained in previous work \cite{macedoElectromagneticApproachCavity2021}:
\begin{ceqn}
\begin{equation}
\label{w_a_b}
    \omega_{a,b}=\frac{1}{2}\left[\omega_c + \omega_0 \pm \sqrt{(\omega_c - \omega_0)^2 + 2 \omega_c \omega_m\frac{W_p}{W_c}}\right],
\end{equation}
\end{ceqn}
where, $\omega_a$ and $\omega_b$ are the eigenfrequencies of the cavity-magnon hybrid system. 

The macroscopic coupling strength, $g$, is related to the width of the splitting between the eigenfrequencies at $\omega_0 = \omega_c$, where $2g = |\omega_a-\omega_b|=\Delta\omega$. 
Therefore, we can then take the eigenfrequencies of the system at $\omega_0 = \omega_c$ from Eq.~(\ref{w_a_b}) which take the form:
\begin{ceqn}
\begin{equation}
    \label{ellipticalS11}
    \omega_{a,b} = \omega_c \pm \frac{1}{2} \sqrt{2 \omega_c \omega_m\frac{W_p}{W_c}}
\end{equation}
\end{ceqn}
We can, therefore use this result to calculate the size of the Rabi splitting:
\begin{ceqn}
\begin{equation}
\label{DeltaW}
    \Delta\omega=\omega_a-\omega_b = \sqrt{2\omega_c\omega_m \frac{W_p}{W_c}}.
\end{equation}
\end{ceqn}

%Include the bias field change for perturbation theory here?

\begin{figure}[ht]
\centering
\includegraphics[width=0.9\linewidth]{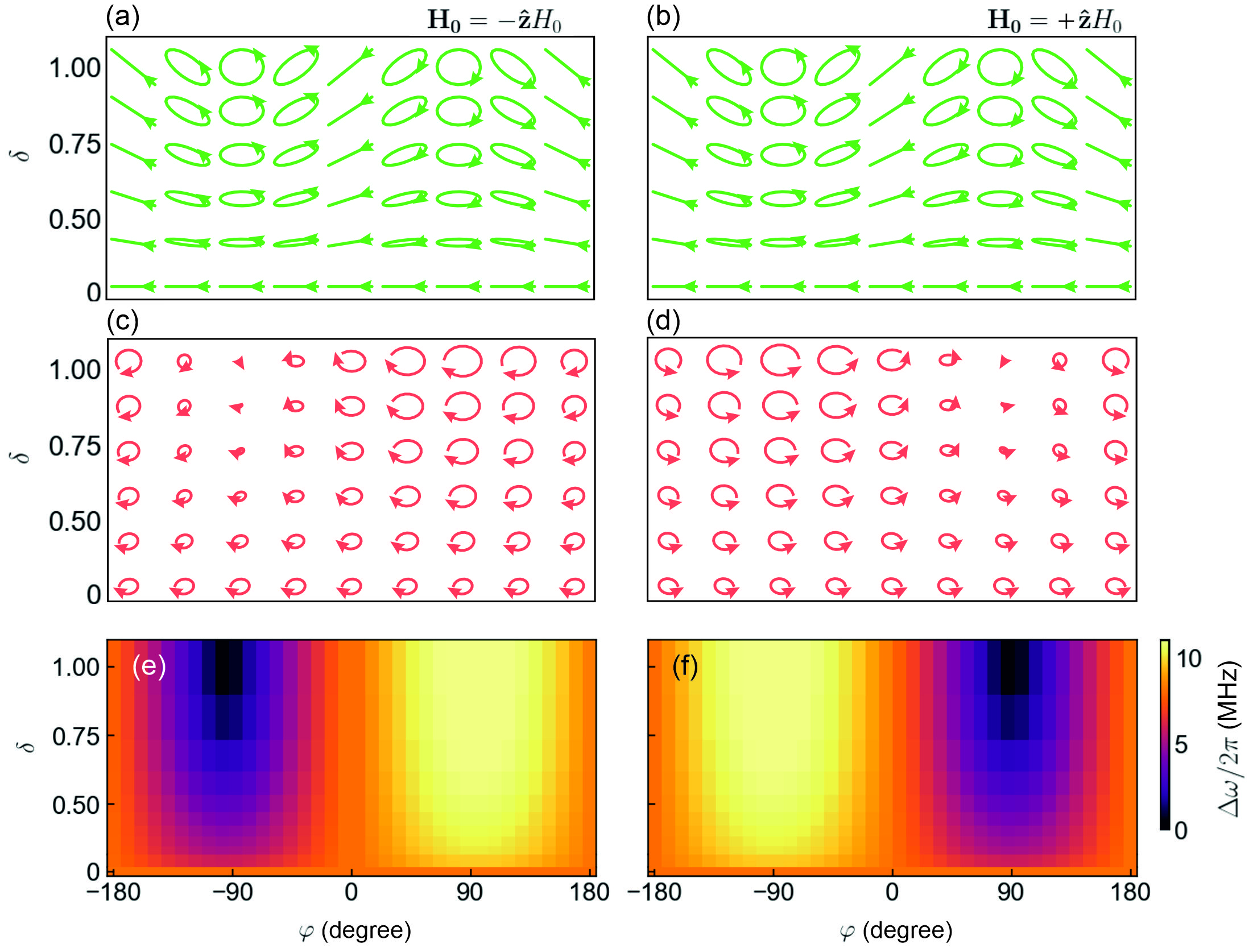}
\caption{(a) and (b) Diagrams illustrating the polarisation of the driving field for different values of $\delta$ and $\varphi$ for a bias field of $+\hat{\mathbf{z}}H_{0}$ and $-\hat{\mathbf{z}}H_{0}$, respectively. (c) and (d) show the subsequent spin precession resulting from the excitation fields and bias fields in (a) and (b), respectively, calculated using the Landau-Lifshitz-Gilbert (LLG) equations. Heat maps summarising the width of the Rabi splitting, $\Delta\omega/2\pi$, as a function of $\delta$ and $\varphi$, calculated using perturbation theory, are presented in (e) for a bias field of $+\hat{\mathbf{z}}H_{0}$ and in (f) for a bias field of $-\hat{\mathbf{z}}H_{0}$.
}
\label{fig:supPerturbationAmpPhase}
\end{figure}

%\noindent The calculated values for $\Delta\omega$ are shown in the following table:

%\begin{table}[h]
%\centering
%\begin{tabular}{|c|c|c|c|}
%\hline
%Excitation Field Polarisation & Perturbation Theory (MHz) & Experimentally Measured (MHz) & Input-Output Theory (MHz) \\
%\hline
%Right Circular & 11.27 & 11.28 & 11.05 \\
%Linear & 7.97 & 7.80 & 7.80 \\
%Left Circular & 0 & 0 & 0 \\
%\hline
%\end{tabular}
%\label{tab:perturbationDeltaW}
%\end{table}

%Quantum - Right 11.045522761381044
%Quantum - Linear 7.803901950975245

%Experiment - Right 11.280000000000179
%Experiment - Linear 7.799999999999585

%Perturbation - Right 11.265788767512293
%Perturbation - Linear 7.967850666127507

Figure \ref{fig:supPerturbationAmpPhase} presents the magnetisation precession dynamics and the hybridisation behaviour observed across the entire range of excitation field polarisations and bias field orientations. The green arrows in panels (a) and (b) depict the polarisation state of the excitation field for various values of $\delta$ and $\varphi$. The red arrows in (c) and (d) correspond to the subsequent magnetisation precession in the YIG sample, calculated from the Landau-Lifshitz-Gilbert (LLG) equations, for bias fields along $+\hat{\mathbf{z}}$ in (c) and $-\hat{\mathbf{z}}$ in (d), respectively.  The calculated values for $\Delta\omega/2\pi$ from Eq. (\ref{DeltaW}) are presented in the heat maps of Fig. \ref{fig:supPerturbationAmpPhase} for a bias field of $+\hat{\mathbf{z}}H_{0}$ in (e) and $-\hat{\mathbf{z}}H_{0}$ in (f).
We see good agreement between these theoretical predictions and the experimental data presented in Fig. \ref{fig:AmpPhase}. 

\subsection*{Special case: Circularly polarised excitation vector field}
$ $

Now if we introduce a left or right circularly polarised excitation vector field at the sample position by plugging $\delta = 1, \varphi=\pm90$ into Eq.~(\ref{eq:pertBeforeRCP}) we obtain: %i.e. for $\mathbf{h^\mp}$:

\begin{ceqn}
\begin{equation}
\frac{\omega - \omega_c}{\omega_c} =-2\Big[\chi_{a} \mp \chi_{b}\Big]\frac{ \displaystyle\int_{\delta v}|h|^2\mathrm{d}v}{2\displaystyle\int_{v}|\mathbf{h_c}|^2 \mathrm{d}v}    
\end{equation}    
\end{ceqn}

\noindent which can be rewritten as:

\begin{ceqn}
\begin{equation}
\frac{\omega - \omega_c}{\omega_c} =-2 \Bigg[\frac{\omega_0\omega_m}{\omega_0^2-\omega^2} \mp \frac{\omega\omega_m}{\omega_0^2-\omega^2}\Bigg]\frac{ \displaystyle\int_{\delta v}|h|^2\mathrm{d}v}{2\displaystyle\int_{v}|\mathbf{h_c}|^2 \mathrm{d}v} 
 =-\frac{\omega_m}{(\omega_0\pm\omega)}  \frac{ \displaystyle\int_{\delta v}|h|^2\mathrm{d}v}{\displaystyle\int_{v}|\mathbf{h_c}|^2 \mathrm{d}v}.   
\end{equation}    
\end{ceqn}

This result can also be obtained by directly combining Eq.~(\ref{mcirc}) directly into Eq.~(\ref{eq:perturbationTheorySupplement}) to obtain:

\begin{ceqn}
\begin{equation}
\frac{\omega - \omega_c}{\omega_c} = -\frac{\omega_m}{\omega_0\pm \omega } \frac{ \displaystyle\int_{\delta v}|\mathbf{h^\mp}|^2\mathrm{d}v}{2\displaystyle\int_{v}|\mathbf{h_c}|^2 \mathrm{d}v}.    
\end{equation}    
\end{ceqn}

\newpage
\section*{Supplementary Note B: Input-output theory of the cavity}

In this section, we delve into the interplay between the incoming and outgoing travelling wave excitations on a cavity, which can be used to find the scattering matrix elements for reflection and transmission coefficients. The cavity modes interact with the continuum of electromagnetic modes outside the cavity. The interaction between the cavity mode $a$ and the continuum can be modelled as \cite{walls2008input}
\begin{ceqn}
\begin{equation}\label{eq8}
    \frac{H}{\hslash}=\omega_a a^{\dagger}a+\int_{-\infty}^{\infty} d\omega[\omega b^{\dagger}(\omega)b(\omega)]+i\int_{-\infty}^{\infty}d\omega g(\omega)[a^\dagger b(\omega)-ab^{\dagger}(\omega)],
\end{equation}
\end{ceqn}
where $b(\omega)(b^{\dagger}(\omega))$ is the annihilation (creation) operator of the continuum of electromagnetic modes satisfying $[b(\omega), b^{\dagger}(\omega^\prime)]=\delta(\omega-\omega^\prime)$, while $g(\omega)$ parameterizes strength of interaction between the cavity mode and the continuum. The Heisenberg equation of motion (EOM) of the mode $a$ is given by
\begin{ceqn}
\begin{align}\label{eq9}
    \dot{a}=-i\omega a-\int_{-\infty}^{\infty}g(\omega)e^{-i\omega(t-t^\prime)}b(t^\prime)dt^\prime.
\end{align}
\end{ceqn}
By the same token, the EOM for the $b$ modes can be obtained, upon formal integration, as
\begin{ceqn}
\begin{align}\label{eq10}
    b(\omega,t)=e^{-i\omega(t-t_{-\infty})}b(\omega,t_{-\infty})+g(\omega)\int_{t_{-\infty}}^te^{-i\omega(t-t^\prime)}a(t^\prime)dt^\prime,
\end{align}
\end{ceqn}
for $t>t_-\infty$ and
\begin{ceqn}
\begin{align}\label{eq11}
    b(\omega,t)=e^{-i\omega(t-t_\infty)}b(\omega,t_\infty)-g(\omega)\int_{t}^{t_\infty}e^{-i\omega(t-t^\prime)}a(t^\prime)dt^\prime,
\end{align}
\end{ceqn}
for $t<t_\infty$. Substituting Eq. (\ref{eq10}) and Eq. (\ref{eq11}) into Eq. (\ref{eq9}) and defining
\begin{ceqn}
\begin{align}\label{eq12}
    a_{in}(t)=-\frac{1}{\sqrt{2\pi}}\int_{-\infty}^{\infty}d\omega e^{-i\omega (t-t_{-\infty})}b({\omega,t_{-\infty}}),\nonumber \\
    a_{out}(t)=\frac{1}{\sqrt{2\pi}}\int_{-\infty}^{\infty}d\omega e^{-i\omega (t-t_{\infty})}b({\omega,t_\infty}),
\end{align}
\end{ceqn}
we obtain
\begin{ceqn}
\begin{align}\label{eq13}
    \dot{a}=-i\omega a-\kappa a+\sqrt{2\kappa}a_{in}(t),  \\
     \dot{a}=-i\omega a-\kappa a-\sqrt{2\kappa}a_{out}(t), \label{eq13.1}
\end{align}
\end{ceqn}
where we have used the Markov approximation $g^2(\omega)=\kappa/\pi$. It follows from Eq. (\ref{eq13}) that 
\begin{ceqn}
\begin{align}\label{eq14}
    a_{in}+a_{out}=\sqrt{2\kappa}a.
\end{align}  
\end{ceqn}
The Eq. (\ref{eq14}) is identical to the Eq. (\ref{eq14_main}) in the main text.

\newpage
\section*{Supplementary Note C: Phase Data}

\begin{figure}[H]
\centering
\includegraphics[width=0.7\linewidth]{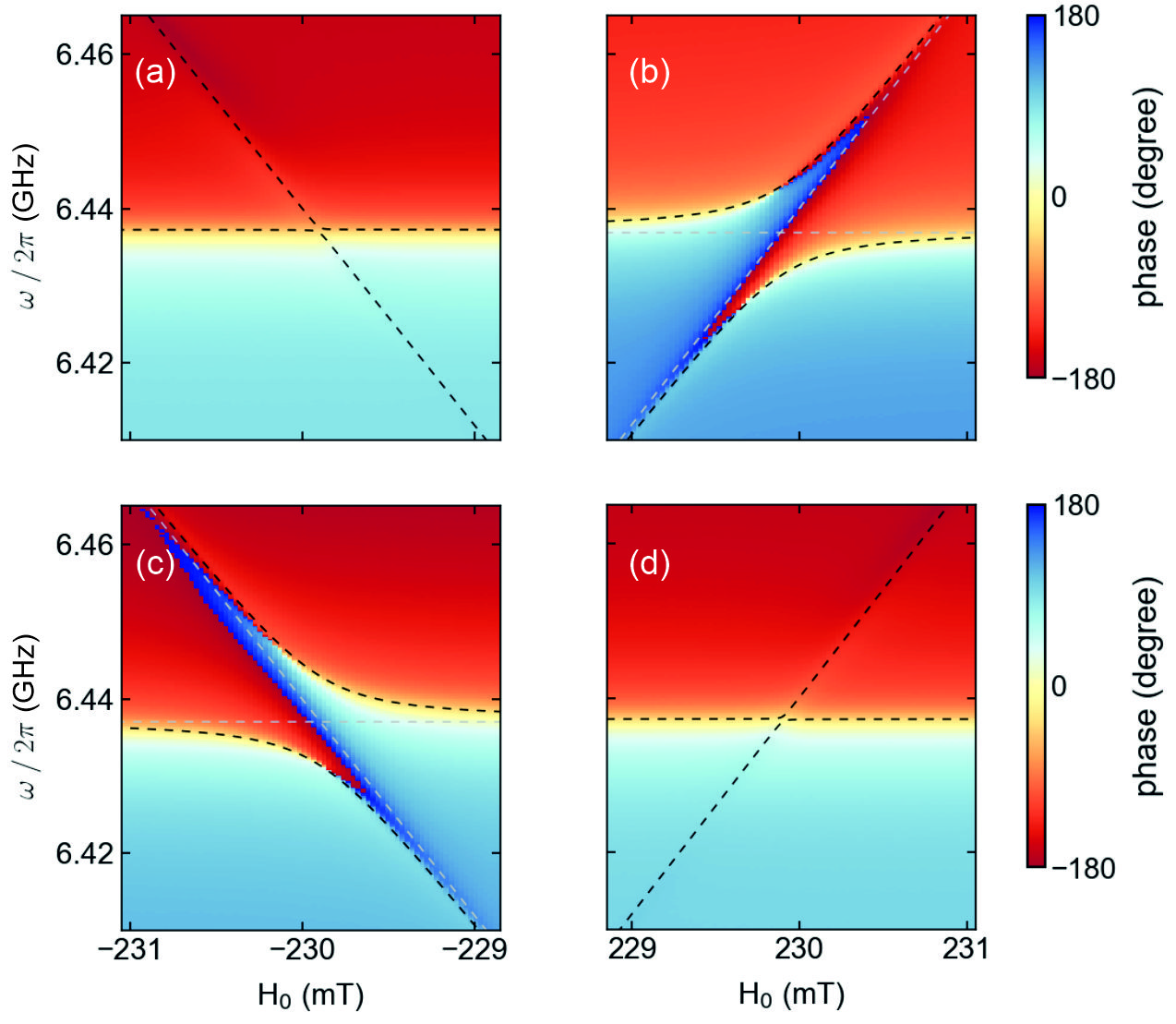}
\caption{
Experimental $S_{11}$ phase spectra and perturbation theory (dashed black lines) of the Rabi splitting close to $\omega_0 = \omega_c$ for (a) $\mathbf{H_{eff}} = -\hat{\mathbf{z}}H_{0}+\mathbf{h}^+$; (b) $\mathbf{H_{eff}} = \hat{\mathbf{z}}H_{0}+\mathbf{h}^+$; (c) $\mathbf{H_{eff}} = -\hat{\mathbf{z}}H_{0}+\mathbf{h}^-$ and (d) $\mathbf{H_{eff}} = \hat{\mathbf{z}}H_{0}+\mathbf{h}^-$. 
The dashed grey lines show the cavity mode and Kittel mode.
}
\label{fig:CircPolPhase}
\end{figure}

The $S_{11}$ parameter is a complex quantity, so an alternative approach to determining the coupling from the amplitude data, as done in Fig. \ref{fig:CircPol}, is to infer the same information from the phase data which is given in Fig.~\ref{fig:CircPolPhase}. % of the $S_{11}$ parameter.
We can see that the phase closely matched the data shown in the main article Fig.~\ref{fig:CircPol} where a sharp change in phase corresponds to a dip in amplitude. 
Thus, two branches are seen for the cases where the chirality of the polarised excitation vector field matches that of the magnon modes (the distance between both branches of phase change at $\omega_c=\omega_0$ can be used to obtain $\Delta\omega$ and therefore $g$) and a single, flat region of sharp change in phase for the cases where the chirality of the magnon mode is opposite to that of the excitation polarisation corresponding to the frequency of the cavity mode $\omega_c$.

\newpage
\section*{Supplementary Note D: Experimental set-up optimisation and calibration}

\subsection*{IQ Mixer -- Calibration}
\begin{figure}[ht]
\centering
\includegraphics[width=0.95\linewidth]{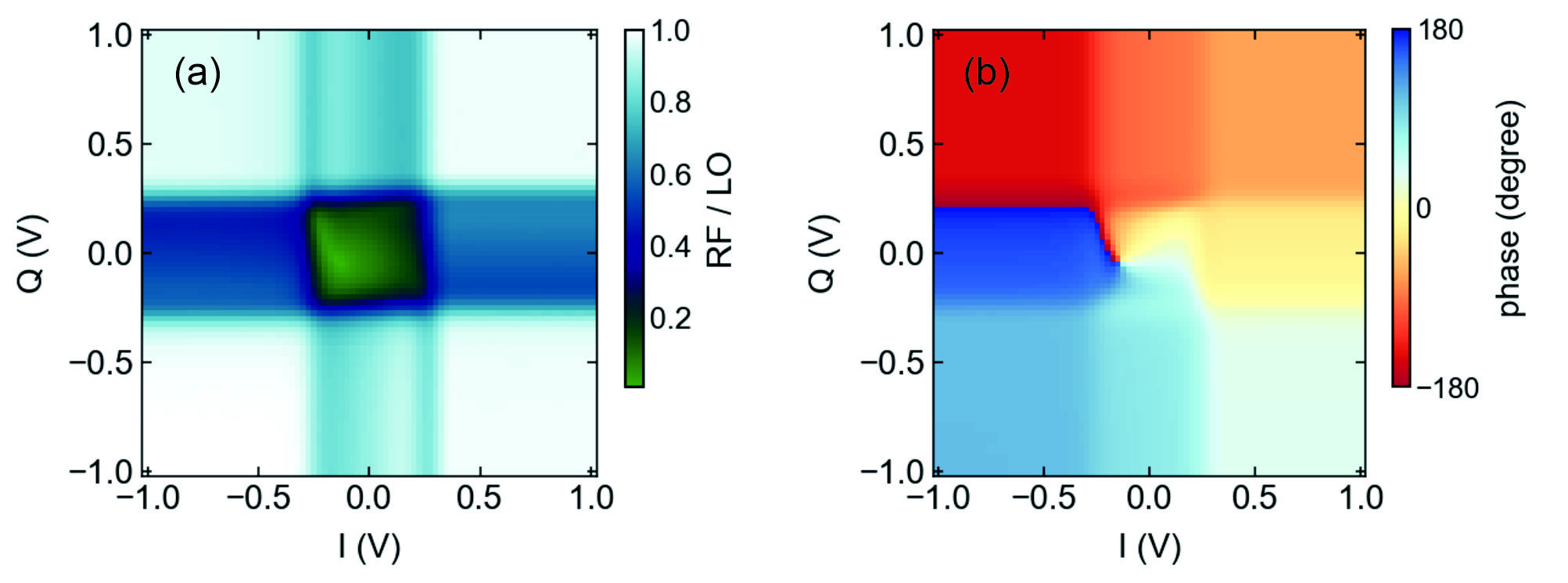}
\caption{IQ mixer (a) amplitude response and (b) phase response to DC voltage input
}
\label{fig:SupIQCalibration}
\end{figure}

To control the phase difference between the two input signals using the IQ mixer, a map of signal phase and amplitude dependence on DC voltage was generated. The phase at ports 1 and 2 was measured and the maps shown in Fig. \ref{fig:SupIQCalibration} were then used to set the phase at port 2 to the desired value. The phase that corresponded to the least attenuation of the signal was selected, i.e. RF/LO $\approx1$.

\subsection*{Extraction of Cavity-Photon and Magnon Dissipation}

\begin{figure}[H]
\centering
\includegraphics[width=0.9\linewidth]{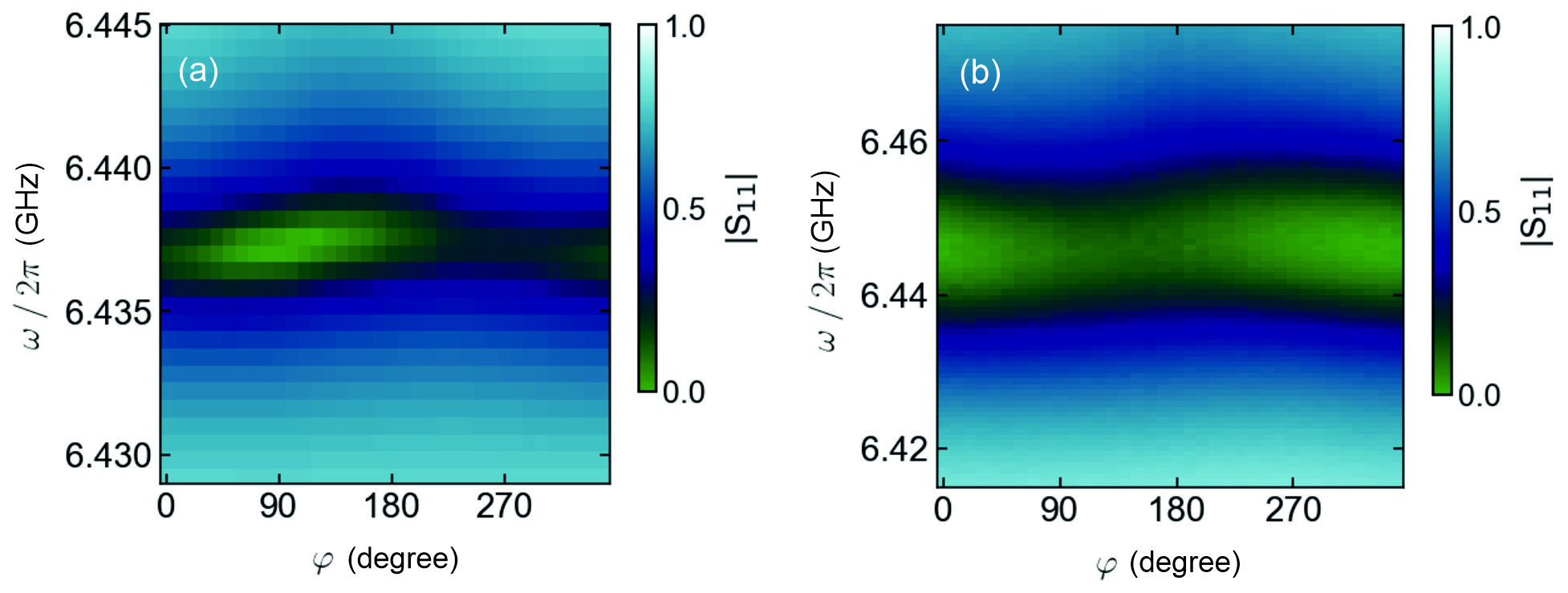}
\caption{Empty cavity mode as a function of phase for the experiment with (a) the 0.25 mm diameter YIG sphere and (b) the 0.5 mm diameter YIG sphere.
}
\label{fig:SupCavityDissipation}
\end{figure}

To determine the cavity dissipation, $\kappa$, a Lorentzian curve was fitted to the resonance curve. The half-width at half-maximum (HWHM) of the Lorentzian was taken as the value of $\kappa$. Given the cavity dimensions, both cavity modes from port 1 and port 2 were theoretically expected to have the same resonance frequency and quality factor. However, due to limitations in the manufacturing process, this was not precisely the case. The cavity was tuned to match the resonance frequencies and quality factors of both modes as closely as possible, but slight deviations remained, resulting in minor changes in the cavity resonance and quality factor as the phase on port 2 was varied. This behaviour is shown in Fig.~\ref{fig:SupCavityDissipation}. To determine the overall cavity resonance $\omega_c$ and dissipation $\kappa$, an average was taken over all phases.

The magnon dissipation, or relaxation rate $\eta$, was determined by fitting a Lorentzian curve to the magnon mode far from the cavity resonance. The value of $\eta$ was calculated from the HWHM, or linewidth $\Delta H$, of the fitted Lorentzian using the expression using the equation $\eta = \gamma \Delta H$, where $\gamma$ is the gyromagnetic ratio \cite{Rezende2020}.

\subsection*{Quality Factor}

The number of photons in a cavity can be estimated using the following equation \cite{Baity2024}:

$$\langle n_{ph} \rangle = \frac{4 P_{in}}{\hslash \omega_c^2} \frac{Q_i^2 Q_c}{(Q_i + Q_c)^2}$$

In our experimental setup, we use two cavity modes to generate excitation fields: TE$_{120}$ from port 1 and TE$_{210}$ from port 2. Thus, we can evaluate the number of photons from each mode. To generate circularly polarised light, the number of photons from each cavity mode must be equal. To achieve this, the coupling quality factor must be equal, since all other variables in the equation are constant for each mode.

$$\frac{\langle n_{ph_1} \rangle}{\langle n_{ph_2} \rangle} = 1 \implies Q_{c1} = Q_{c2}$$

The coupling strength of the antenna to the cavity mode was adjusted until the coupling quality factor for each mode matched as closely as possible.

\newpage
\section*{Supplementary Note E: Superposition of Two Cavity Fields}

\begin{figure}[ht]
\centering
\includegraphics[width=0.95\linewidth]{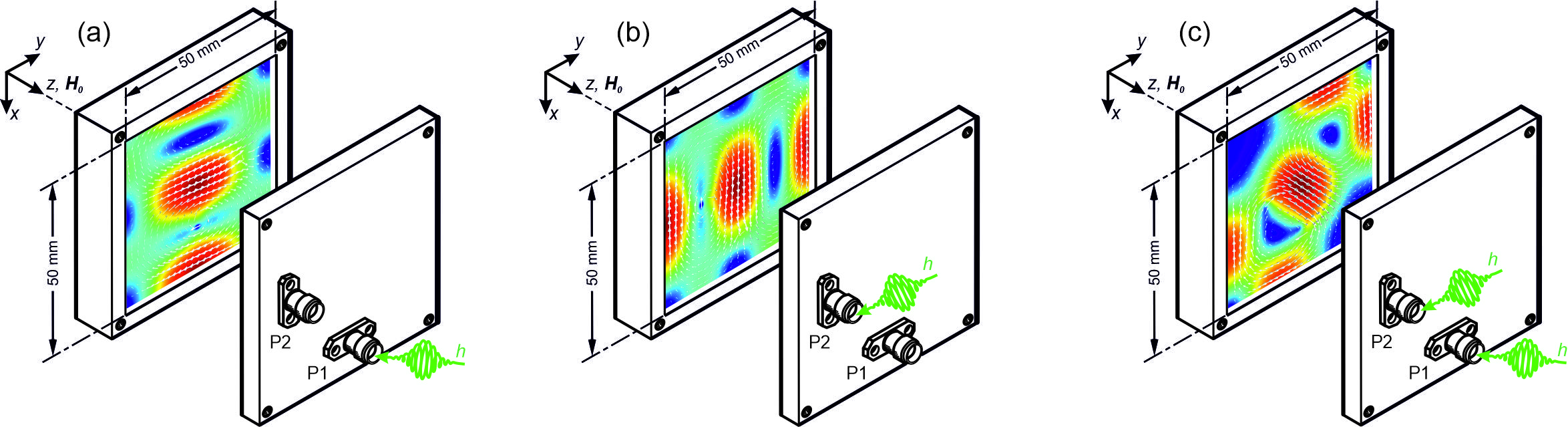}
\caption{The cross-sectional field configuration at $z$ = 2.5~mm generated through capacitive coupling (simulated with COMSOL) within a square cavity resonator. 
(a) The oscillating magnetic field when excited only through port 1 (P1), (b) the oscillating magnetic field when excited only through port 2 (P2) and (c) the oscillating magnetic field when excited equally through both port 1 (P1) and port 2 (P2).
}
\label{fig:Field_g_Position}
\end{figure}

The TE$_{120}$ mode has the field profile shown in Fig.~\ref{fig:Field_g_Position}(a) and was obtained by exciting port 1 of the cavity. 
The oscillating $\mathbf{h_c}$ intensity profile shown has an anti-node at the centre.
At this central point, the $y-$component of the magnetic field, $h_{cy}$, is zero. Thus, we can neglect $h_{cy}$, and consider the excitation vector field to be linearly polarised in the $x-$direction. 
Port 2 generates an identical mode to port 1, however, it is placed in such a position that the fields it generates are rotated by 90$^\circ$, relative to the first port -- i.e, the TE$_{210}$ mode, as shown in panel (b). This mode only has a y-component at the centre of the cavity. 
When exciting with both ports with equal excitation, as shown in panel (c), the cavity excitation vector fields now have an $x-$component and a $y-$component at the centre. The superposition of these fields still generates a linearly polarised excitation, but the direction of the overall oscillating magnetic field in the centre is now rotated by 45$^\circ$. 
By controlling the phase and amplitude of the signal into the second coupler relative to the signal at the first coupler, the $y-$component of the driving field at the centre can be modified, allowing for more complex driving fields, such as circularly polarised light.

\newpage
\section*{Supplementary Note F: Limitation of Experimental Set Up}

\subsection*{Introduction of transmitted energy}
$ $

At higher amplitude ratios, specifically when $\delta>1$, the experimental setup faced limitations that stopped it from operating correctly. %our measurements deviate from the expected purely reflective character.

\begin{figure}[ht]
\centering
\includegraphics[width=0.95\linewidth]{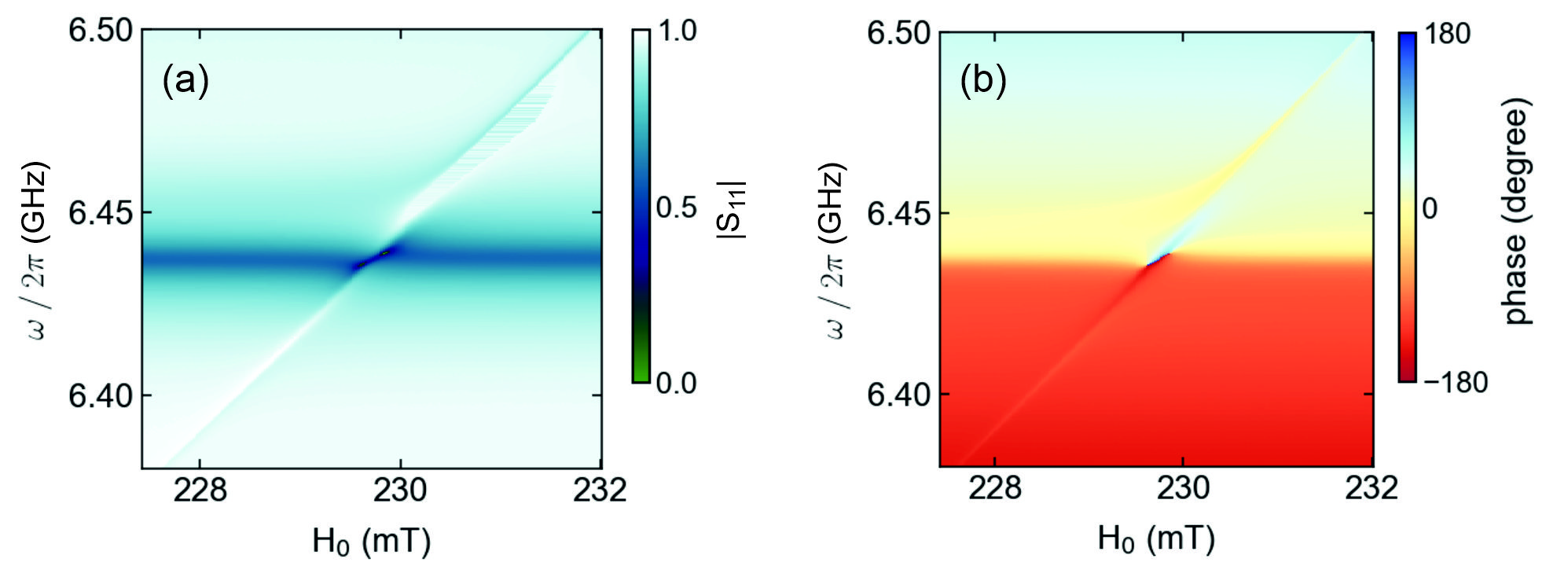}
\caption{Experimentally measured $|S_{11}|$ amplitude data in (a) and phase data in (b) for $\delta\approx1.3$, $\varphi\approx+90^\circ$ and $\mathbf{H_0}=+\mathbf{\hat{z}}H_0$}
\label{fig:SupLevelAttraction1}
\end{figure}

The reflection data presented in Fig.~\ref{fig:SupLevelAttraction1} corresponds to measurements obtained at approximately $\delta\approx1.3$ and $\varphi\approx+90$. The observed spectrum resembles level attraction% (reported in previous two-tone experiments \cite{boventer19a, harder18, gardin2024})
, however, it is unlikely to represent a true manifestation of this phenomenon. According to perturbation theory, level attraction is not expected to occur for any polarisation state. This is also in agreement with our quantised in-and-out model by the absence of non-hermitian terms. Instead, we interpret this observed behavior as an artefact stemming from our experimental setup.

While our experimental setup is designed to measure reflection from port 1, an amplitude ratio of $\delta>1$ results in a higher excitation at port 2 in comparison to port 1. At these amplitude ratios, energy is transmitted from port 2 to port 1 and subsequently, the experimental conditions can no longer be considered purely reflective, but begin to include a transmission component. To confirm, we increased the amplitude ratio, eventually observing a transition in the spectrum to a pure transmission profile. Thus, we believe the appearance resembling level attraction is likely a result of the combination of reflection and transmission in the spectrum rather than an authentic manifestation of level attraction.

Consequently, this experiment is limited to $\delta \leq 1$. However, in order to investigate $\delta\geq 1$, one can simply place the circulator on the other signal path to measure the reflection from port 2 of the cavity (into port 2 of the VNA)
instead. In this configuration, when port 2 exhibits higher excitation compared to port 1, i.e. when $\delta>1$, we can still obtain a purely reflection measurement. We expect the same hybridisation behaviour for $0 \leq \delta \leq 1$ as for $1 \leq \delta < \infty$, since the excitation polarisation for both these ranges is identical. For this reason, we limited our investigation to varying $\delta$ from 1 to 0.

\newpage
\subsection*{Size Effects}
$ $

\begin{figure}[ht]
\centering
\includegraphics[width=0.95\linewidth]{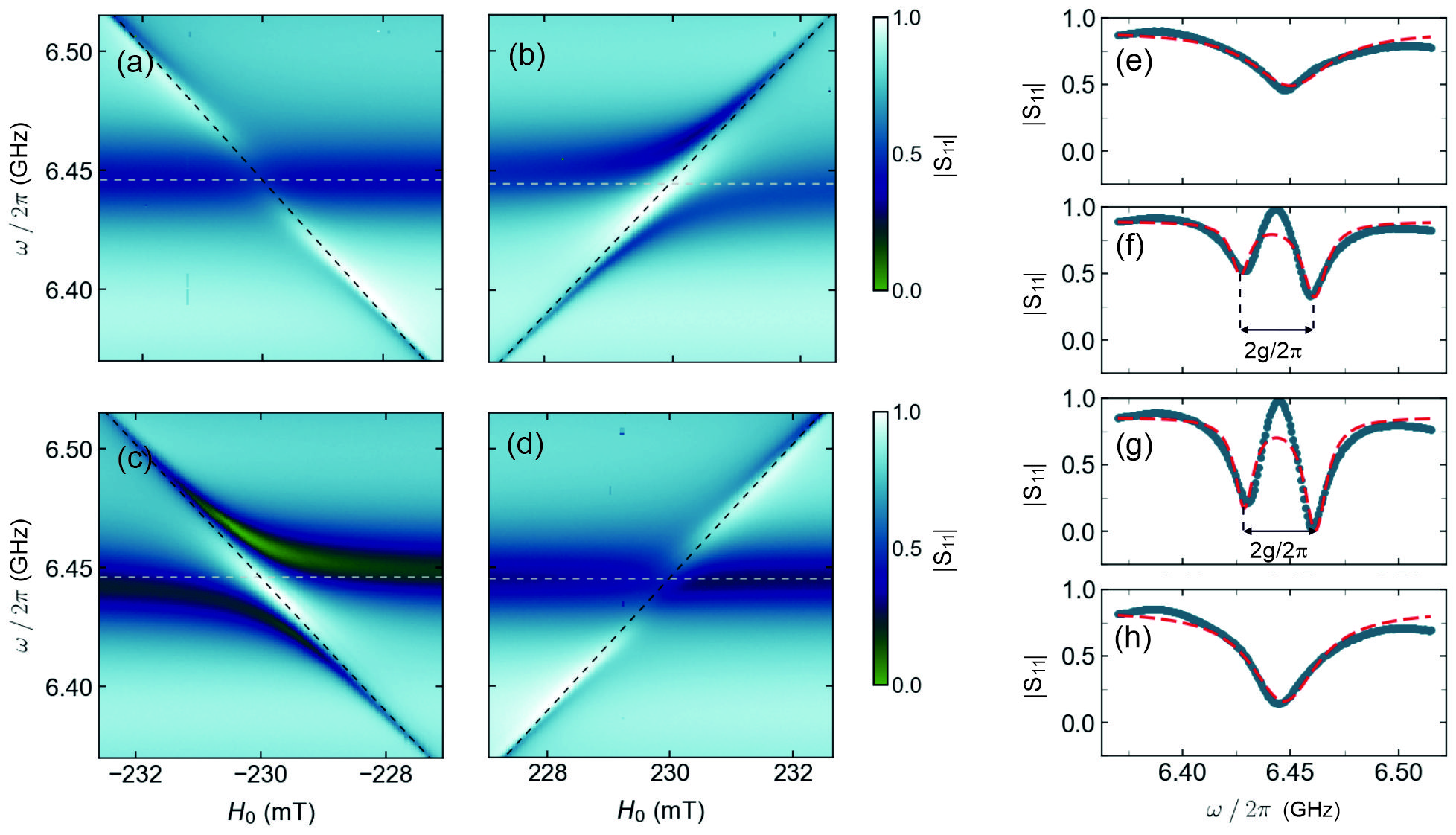}
\caption{%Dont seem to mention YIG in first part
Experimental spectra of the Rabi splitting using a 0.5mm diameter YIG sphere for (a)~$\mathbf{H_{eff}} = -\hat{\mathbf{z}}H_{0}+\mathbf{h}^+$; (b)~$\mathbf{H_{eff}} = \hat{\mathbf{z}}H_{0}+\mathbf{h}^+$; (c)~$\mathbf{H_{eff}} = -\hat{\mathbf{z}}H_{0}+\mathbf{h}^-$ and (d)~$\mathbf{H_{eff}} = \hat{\mathbf{z}}H_{0}+\mathbf{h}^-$. 
The dashed lines show the cavity mode (grey) and Kittel mode (black). 
(e)-(h) show the corresponding $|S_{11}|$ parameter measured when $\omega_0 = \omega_c$.
}
\label{fig:Sup0.5res1}
\end{figure}

\begin{figure}[ht]
\centering
\includegraphics[width=0.95\linewidth]{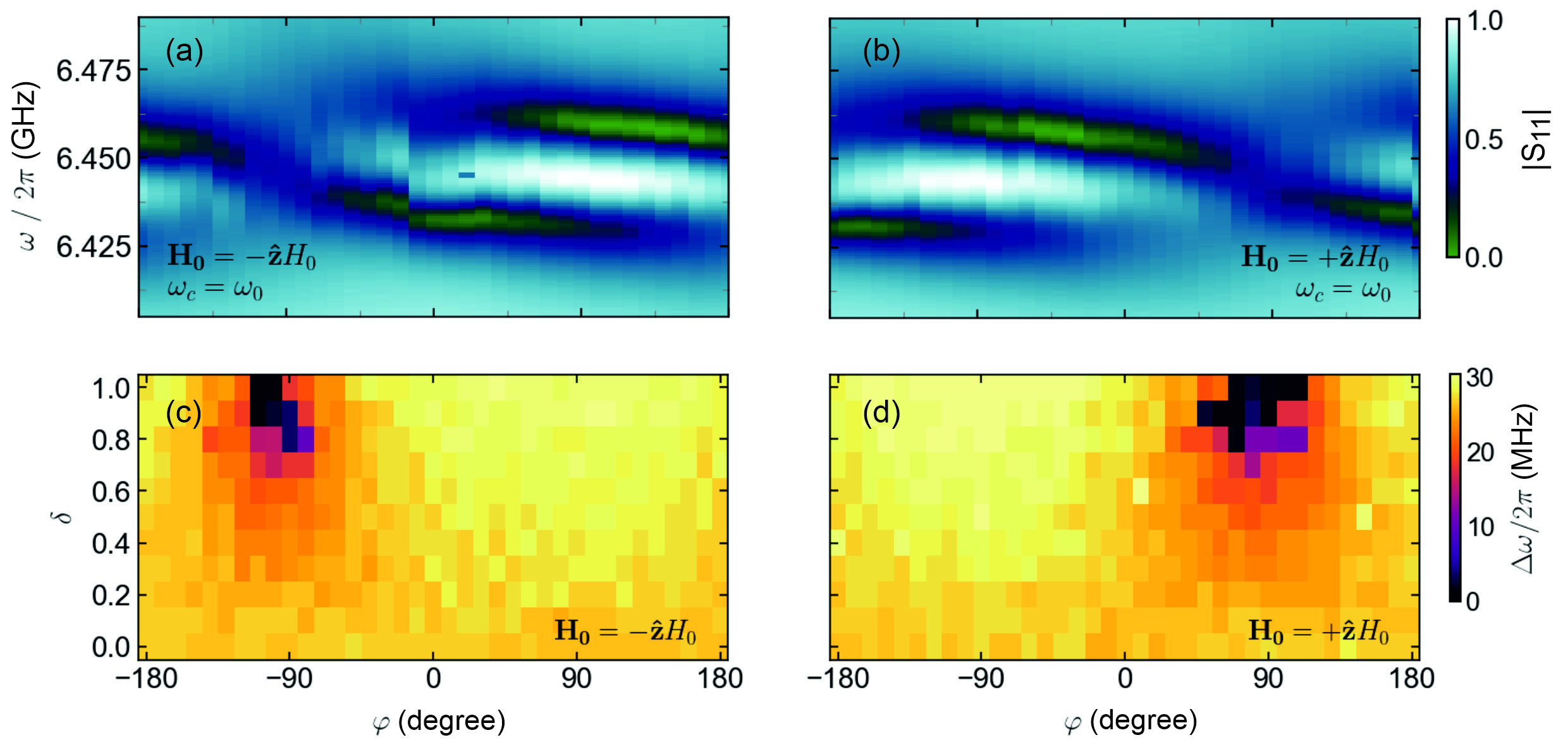}
\caption{
The experimental amplitudes of the $|S_{11}|$ parameter for various 
values of $\varphi$ when $\omega_0 = \omega_c$ using a 0.5mm diameter YIG sphere is shown for a bias field of $+\hat{\mathbf{z}}H_{0}$ in (a) and $-\hat{\mathbf{z}}H_{0}$ in (b). 
The experimentally measured Rabi splitting, $\Delta\omega/2\pi$, for various $\delta$ and $\varphi$ using a 0.5mm diameter YIG sphere for the same bias fields are shown in (c) and (d) respectively.
}
\label{fig:Sup0.5Posres2}
\end{figure}

The measurements were repeated for larger samples; %, but the results were not as satisfactory, although they partially worked. The 
results for the 0.5 mm diameter YIG sphere are shown in Figs.~\ref{fig:Sup0.5res1}--\ref{fig:Sup0.5Posres2}. From Fig.~\ref{fig:Sup0.5res1}, it is evident that with right and left circularly polarised excitation, we were still able to sweep across coupling regimes---level repulsion and annihilation of the cavity magnon-polaritons. Similar to the case of the 0.25 mm diameter YIG sphere, we have field non-reciprocity, so this system could still find applications in information processing, such as an on/off switch.

%A summary of the effect of changing phases is given in Fig.~\ref{fig:ElliptPol}. 
Fig.~\ref{fig:Sup0.5Posres2} summarises the hybridisation behaviour under various driving conditions. Panel (a) and (b) show the effect changing the phases on port 2 relative to port 1 with an amplitude ratio of $\delta=1$ has on the hybridisation behaviour. In comparison to Fig. \ref{fig:ElliptPol} (the results for the 0.25 mm diameter YIG sphere), we can observe that the path loosely follows the same shape. We still see a mode crossing and mode repulsion for left- and right circularly polarised fields, with respect to the bias field $\mathbf{H_0}$, consistent with the smaller sample. However, when exciting with fields near left circular polarisation, the shape begins to deviate from the expected form. We believe this artefact emerges as a result of the larger sample. Since it is larger, the sample no longer interacts with the superposition of cavity modes as a single mode but rather as two degenerate modes. As a result, the sample was coupling more strongly to one mode compared to the other, leading to the observed shape distortion.

Panels (c) and (d) can be compared to Fig.~\ref{fig:AmpPhase}. Again, the shape is somewhat similar but not exactly the same. The most noticeable difference was that the enhanced coupling region (near the right circularly polarised excitation) was broader, and the area showing decreased coupling (near the left circularly polarised excitation) was narrower. Additionally, the enhancement in coupling for right-circularly polarised excitation compared to linearly polarised excitation was not as significant as observed with the smaller sample. With the smaller sample, we saw an enhancement in the coupling strength by a factor of $\sqrt{2}$, whereas for the 0.5 mm diameter YIG sphere, the enhancement in coupling strength was not as significant (increased from 12.5 MHz when exciting the sample with a single tone, for $\delta=0$, to 15.5 MHz for right-circularly polarised excitation). 

The experimental results obtained using the 1mm diameter YIG sphere suggested that the sample was no longer coupled to a single superimposed tone. Instead, it exhibited distinct coupling behaviours for each individual tone. Notably, for left circularly polarised excitation fields with respect to the bias field, we observed a level crossing. However, the spectra also displayed a component of level repulsion. This type of spectral behaviour, featuring both level crossing and level repulsion, has been seen previously in the literature \cite{Rao2019, Yu2020, Bourhill2023}. The datasets for all sample sizes is presented below.

\begin{figure}
\centering
\includegraphics[width=0.95\linewidth]{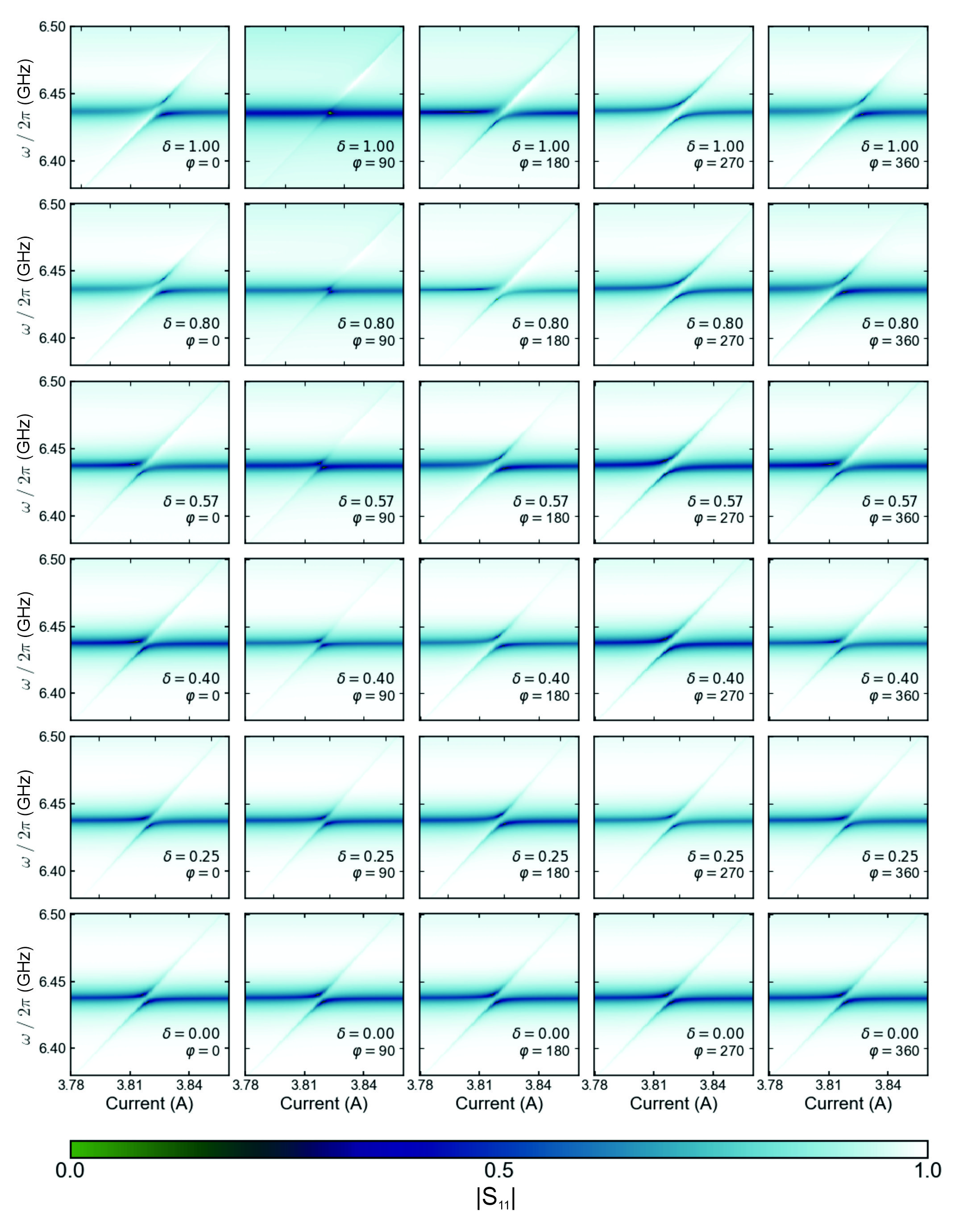}
\caption{Experimental data using 0.25mm diameter YIG sphere for $\mathbf{H_0}=+\hat{\mathbf{z}}H_{0}$. 
}
\label{fig:Sup0.25Pos}
\end{figure}

\begin{figure}
\centering
\includegraphics[width=0.95\linewidth]{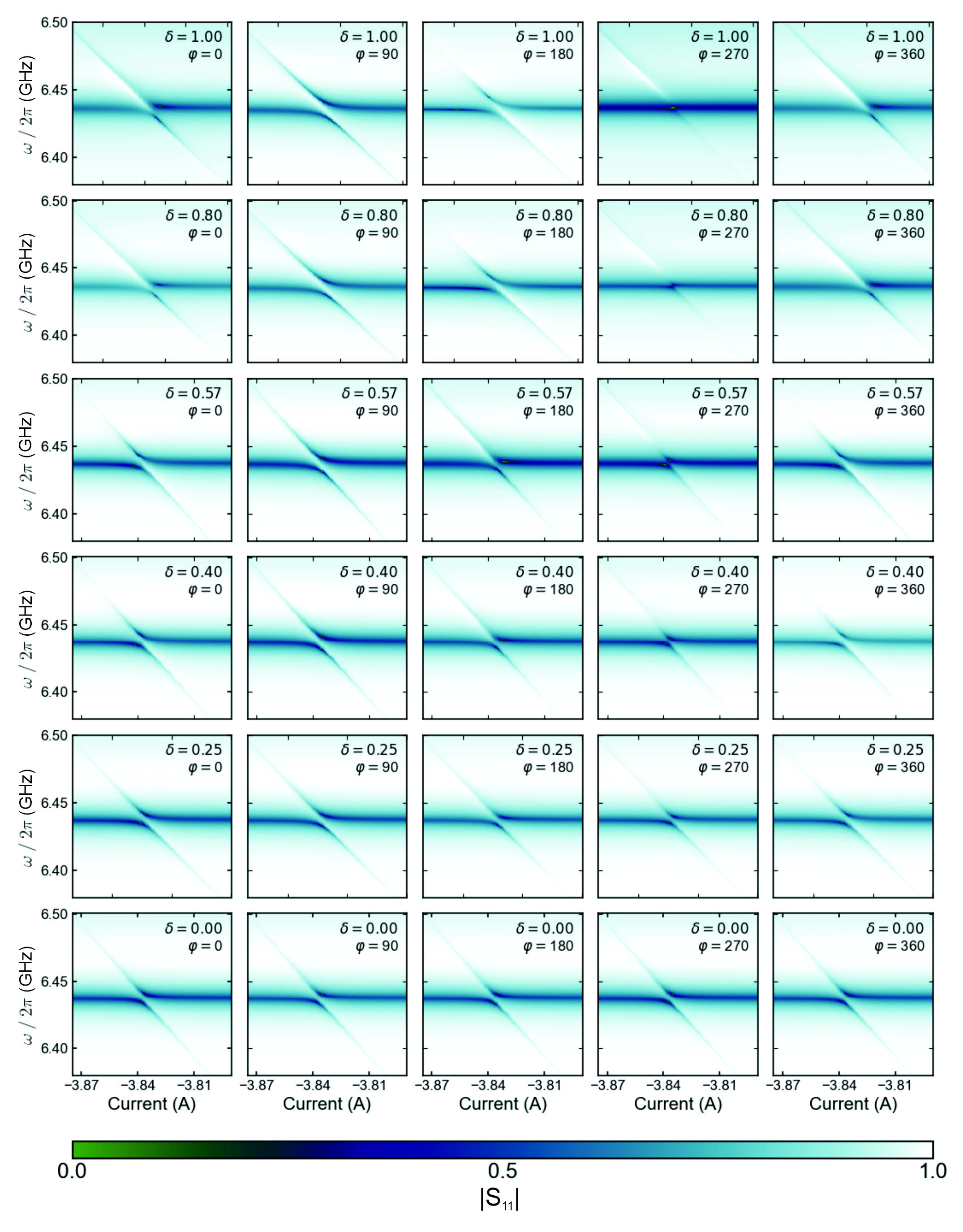}
\caption{Experimental data using 0.25mm diameter YIG sphere for $\mathbf{H_0}=-\hat{\mathbf{z}}H_{0}$. 
}
\label{fig:Sup0.25Neg}
\end{figure}

\begin{figure}
\centering
\includegraphics[width=0.95\linewidth]{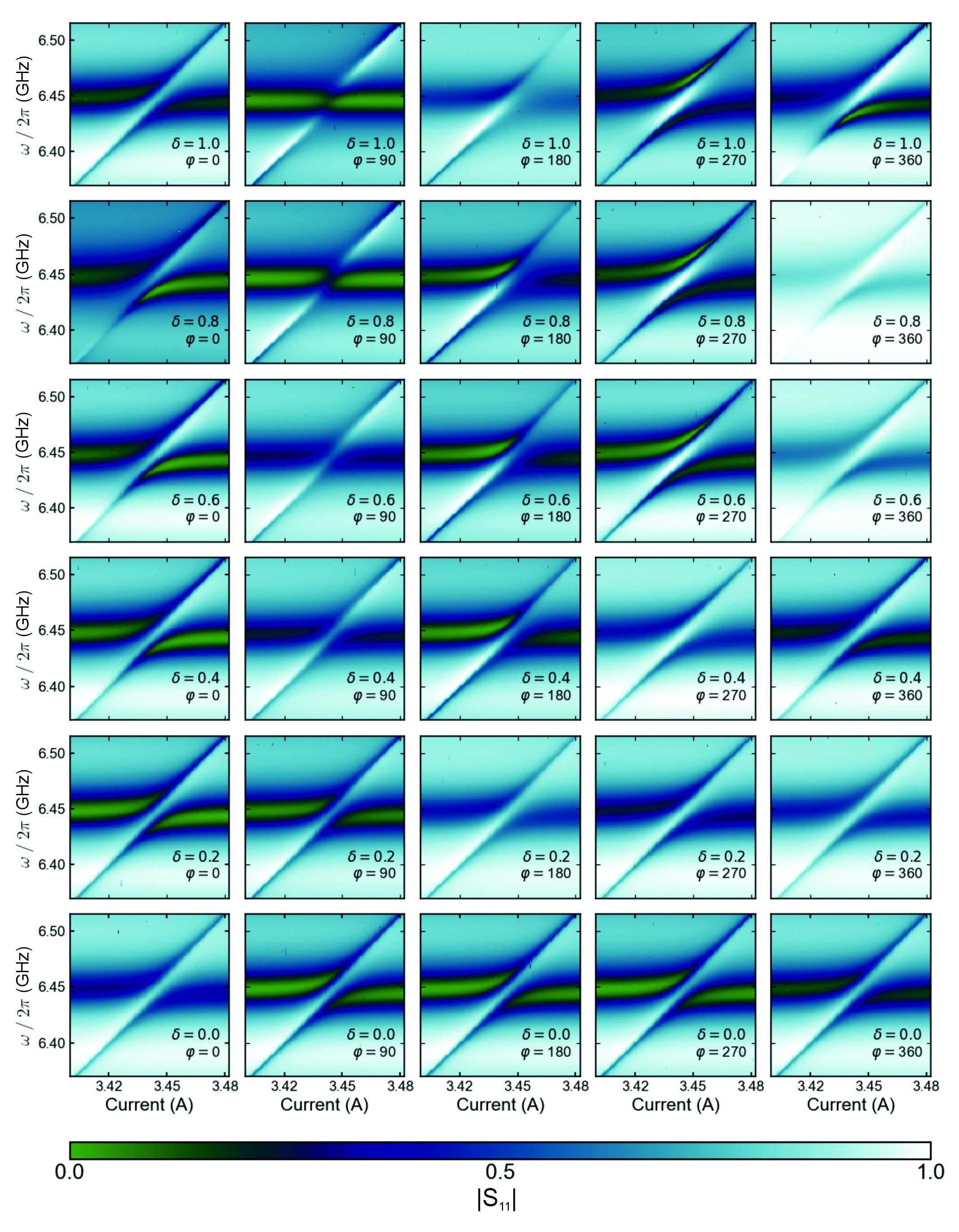}
\caption{Experimental data using 0.5mm diameter YIG sphere for $\mathbf{H_0}=+\hat{\mathbf{z}}H_{0}$
}
\label{fig:Sup0.5Pos}
\end{figure}

\begin{figure}
\centering
\includegraphics[width=0.95\linewidth]{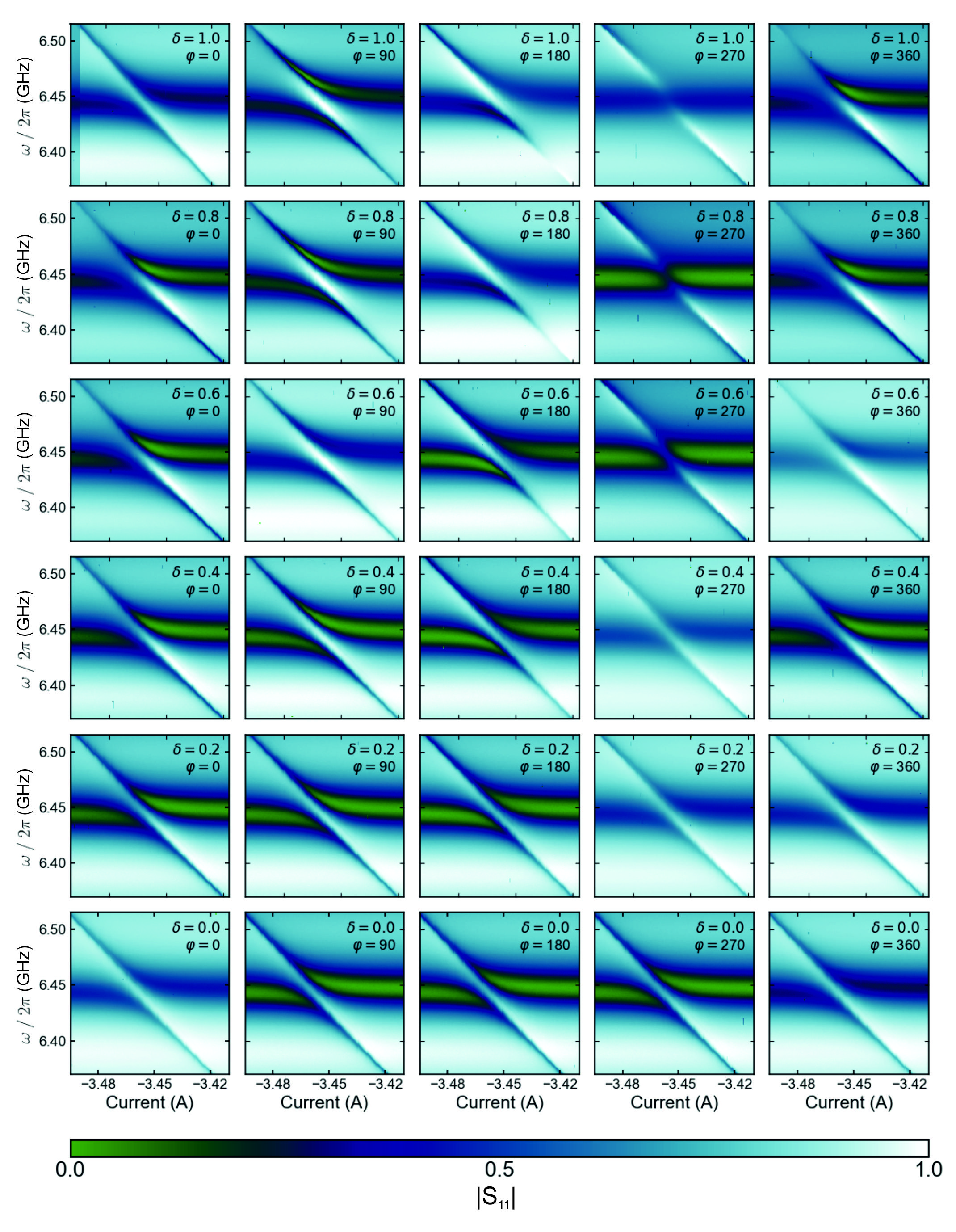}
\caption{Experimental data using 0.5mm diameter YIG sphere for $\mathbf{H_0}=+\hat{\mathbf{z}}H_{0}$
}
\label{fig:Sup0.5Neg}
\end{figure}

\begin{figure}
\centering
\includegraphics[width=0.95\linewidth]{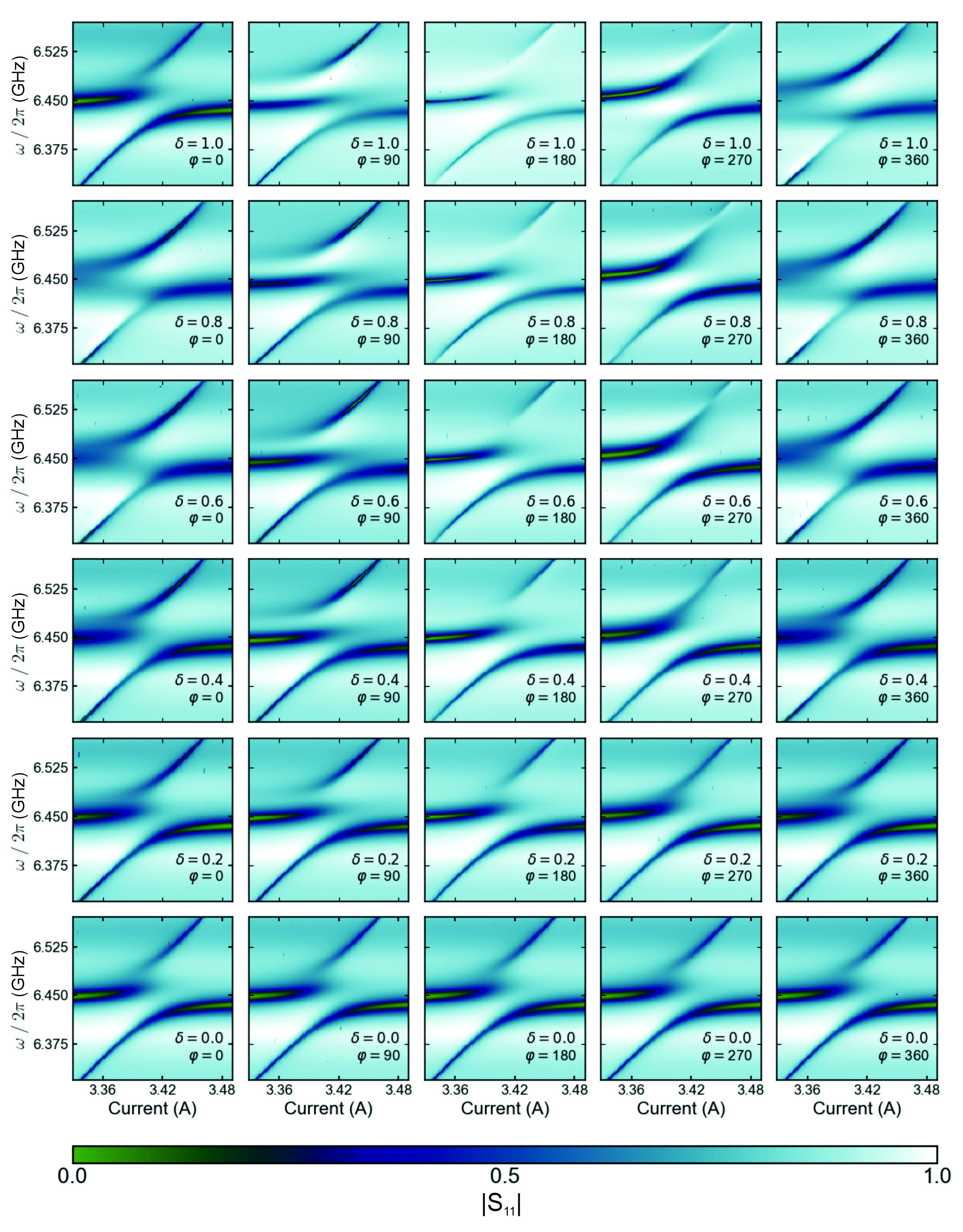}
\caption{Experimental data using 1mm diameter YIG sphere for $\mathbf{H_0}=+\hat{\mathbf{z}}H_{0}$
}
\label{fig:Sup1Pos}
\end{figure}

\begin{figure}
\centering
\includegraphics[width=0.95\linewidth]{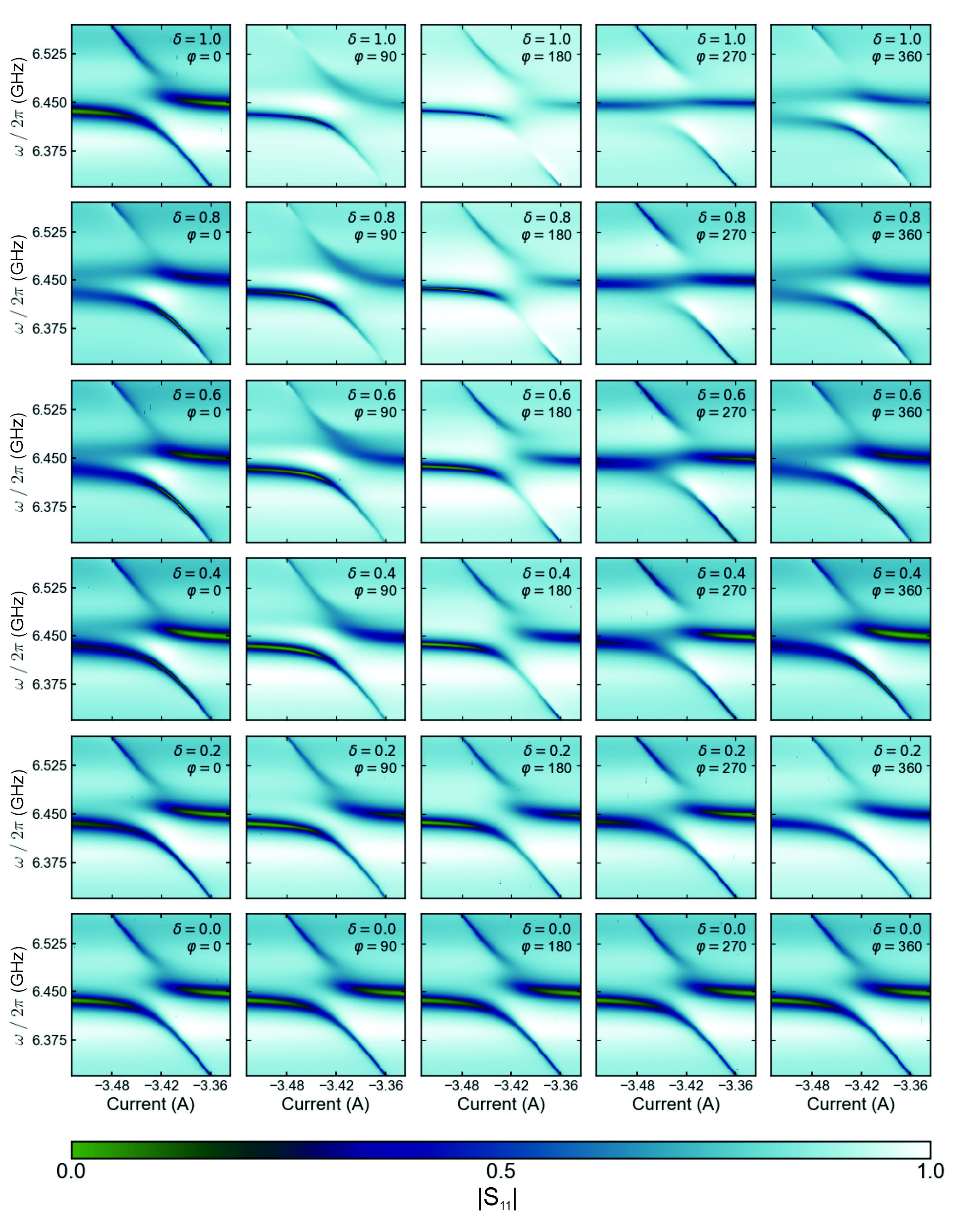}
\caption{Experimental data using 1mm diameter YIG sphere for $\mathbf{H_0}=-\hat{\mathbf{z}}H_{0}$
}
\label{fig:Sup1Neg}
\end{figure}

\end{document}